\documentclass[useAMS,usenatbib]{mn2e}

\usepackage{graphicx}
\usepackage{footmisc}
\usepackage{color} 
\usepackage{rotating}
\usepackage{amssymb}
\usepackage{float}
\usepackage{longtable,lscape}

\newcommand{\lum}{{\rm erg\,s^{-1}}}
\newcommand{\flux}{{\rm erg\,s^{-1}cm^{-2}}}
\newcommand{\kms}{{\rm km\,s^{-1}}}

\newcommand{\aj}{AJ}

\newcommand{\xmm}{\textit{XMM-Newton }}

\newcommand{\mic}{$\mu$m}                      %
\newcommand{\wise}{$\it WISE$}                      %
\newcommand{\swise}{$\it WISE$ }                      %
\title[Mid-IR continuum of AGN]{Revisiting the relationship between
  6\,\mic\, and 2-10 keV continuum luminosities of AGN} \author[Mateos
  et al.] {S. Mateos$^{1}$\thanks{E-mail: mateos@ifca.unican.es},
  F. J. Carrera$^{1}$, A. Alonso-Herrero$^{1}$\thanks{Augusto
    G. Linares Senior Research Fellow}, E. Rovilos$^{2}$,
  A. Hern\'an-Caballero$^{1}$, \newauthor X. Barcons$^{1}$, A. Blain$^{3}$, A. Caccianiga$^{4}$, R. Della Ceca$^{4}$, P. Severgnini$^{4}$
\smallskip \\
\footnotesize
$^{1}$ Instituto de F\'isica de Cantabria (CSIC-Universidad de Cantabria), 39005, Santander, Spain\\
$^{2}$ Institute for Astronomy, Astrophysics, Space Applications \& Remote
Sensing, National Observatory of Athens, 15236 Palaia Penteli, Greece \\
$^{3}$ Physics and Astronomy, University of Leicester, University Road, Leicester LE1 7RH, UK\\
$^{4}$ INAF-Osservatorio Astronomico di Brera, via Brera 28, 20121 Milano, Italy}
\begin{document}

\date{Accepted 2015 February 11. Received 2015 February 10; in
  original form 2014 October 2}

\pagerange{\pageref{firstpage}--\pageref{lastpage}} \pubyear{2012}

\maketitle

\label{firstpage}

\begin{abstract}
We have determined the relation between the AGN luminosities at
rest-frame 6\,\mic\, associated to the dusty torus emission and at
2-10 keV energies using a complete, X-ray flux limited sample of 232
AGN drawn from the Bright Ultra-hard XMM-Newton Survey. The objects
have intrinsic X-ray luminosities between $10^{42}$ and $10^{46}{\rm
  \,erg\,s^{-1}}$ and redshifts from 0.05 to 2.8. The rest-frame
6\,\mic\, luminosities were computed using data from the Wide-Field
Infrared Survey Explorer and are based on a spectral energy
distribution decomposition into AGN and galaxy emission. The best-fit
relationship for the full sample is consistent with being linear,
$L_{\rm 6\,\mu m}$$\propto$$L_{\rm 2-10\,keV}^{0.99\pm0.03}$, with
intrinsic scatter, $\Delta$\,log\,$L_{\rm 6\,\mu m}$$\sim$0.35
dex. The $L_{\rm 6\,\mu m}/L_{\rm 2-10\,keV}$ luminosity ratio is
largely independent on the line-of-sight X-ray absorption. Assuming a
constant X-ray bolometric correction, the fraction of AGN bolometric
luminosity reprocessed in the mid-IR decreases weakly, if at all, with
the AGN luminosity, a finding at odds with simple receding torus
models. Type 2 AGN have redder mid-IR continua at rest-frame
wavelengths $<$12\,\mic\, and are overall $\sim$1.3-2 times fainter at
6 \mic\, than type 1 AGN at a given X-ray luminosity. Regardless of
whether type 1 and type 2 AGN have the same or different nuclear dusty
toroidal structures, our results imply that the AGN emission at
rest-frame 6\,\mic\, is not isotropic due to self-absorption in the
dusty torus, as predicted by AGN torus models. Thus, AGN surveys at
rest-frame $\sim$6\,\mic\, are subject to modest dust obscuration
biases.
\end{abstract}

\begin{keywords}
galaxies: active-quasars: general-infrared: galaxies
\end{keywords}

\section{Introduction}
In the standard unified model of Active Galactic Nuclei (AGN), the
observed differences between AGN sub-classes are solely explained as
the result of an orientation effect with respect to an optically and
geometrically thick distribution of molecular gas and dust on tens of
parsecs scales, frequently assumed to have a toroidal geometry
(\citealt{antonucci93}; \citealt{urry95}). This material, hereafter
referred to as the dusty torus, absorbs the AGN bolometric luminosity
output produced at rest-frame UV/optical and X-ray wavelengths in the
accretion disk/corona and re-emits it mainly in the mid-infrared
(mid-IR) regime (rest-frame wavelengths $\sim$5-30\,\mic). Thus, it is
clear that the mid-IR spectral range offers a unique opportunity to
study the obscured accretion phenomenon. Surveys conducted using data
from Spitzer-IRAC have identified a population of highly obscured AGN
whose number density is comparable to that observed for unobscured
AGN. Many of these objects might be hidden at optical and X-ray
wavelengths by Compton-thick material (line-of-sight rest-frame gas
column densities ${\rm N_H > 1.5\times 10^{24}\,cm^{-2}}$;
\citealt{martinez-sansigre05}).

Uncovering the mid-IR thermal continuum associated with the AGN dusty
torus emission can be a difficult task, due to contamination from the
AGN host galaxies. High (subarcsecond) spatial resolution observations
can isolate the AGN emission from other contaminants on nearby objects
with favourable geometries. Alternatively, mid-IR spectral
decomposition can provide reliable estimates of the AGN dusty torus
contribution to the infrared light (e.g. \citealt{ballo14}). Studies
such as these have found a tight correlation between the intrinsic
luminosity, traced by hard (2-10 keV) X-rays, and reprocessed
mid-infrared emission of AGN over more than three orders of magnitude
in luminosity (\citealt{lutz04}; \citealt{ramos07}; \citealt{horst08};
\citealt{gandhi09}; \citealt{levenson09}; \citealt{asmus11};
\citealt{mason12}; \citealt{asmus14}). The relationship between hard
X-ray and mid-IR luminosities seems to be identical for AGN optically
classified as type 1 and type 2, even for Compton-thick AGN. If the
hard X-ray emission is a good tracer of both the intrinsic power of
AGN and the heating source of the dusty torus, then these studies
provide strong observational support for the mid-IR light being a
reliable proxy for the AGN intrinsic power, regardless of any
obscuration. Unfortunately, the AGN samples involved in these studies
are highly incomplete and biased towards infrared-bright
objects. Recent studies of local AGN detected with the Swift Burst
Alert Telescope and AKARI all-sky surveys have shown a tight
correlation between the hard X-ray ($>$10 keV energies) and
monochromatic mid-IR (9 and 18\,\mic) luminosities and that unabsorbed
and absorbed AGN follow the same correlation (\citealt{ichikawa12};
\citealt{matsuta12}).

The sensitive all-sky infrared survey conducted with the Wide-field
Infrared Survey Explorer (\wise; \citealt{wright10}) at 3.4, 4.6, 12
and 22\,\mic\, will dramatically increase our AGN census over a
significant fraction of the age of the Universe improving greatly our
understanding of the obscured accretion phenomenon in luminous
systems. It has already been demonstrated that mid-IR colour-based
selection techniques with \wise, targeting the characteristic red
power-law thermal continuum emission from the AGN dusty torus
($f_\nu$\,$\propto$\,$\nu^{\alpha}$ with $\alpha$\,$\leq$\,-0.5;
\citealt{alonso-herrero06}), are highly reliable and effective at
uncovering both unobscured and obscured luminous AGN
(e.g. \citealt{assef10}; \citealt{mateos12}; \citealt{stern12};
\citealt{assef13}; \citealt{mateos13}; \citealt{yan13}). As
demonstrated in \citet{mateos13}, \swise can potentially uncover many
Compton-thick AGN missed at other wavelengths, at least up to
$z$$\sim$1. Despite such progress, the physical properties of the AGN
revealed with \swise remain poorly constrained.

In this paper we investigate how well we can isolate the AGN emission
associated to the dusty torus in the mid-IR regime using the \swise
broad band photometric data and, more importantly, whether the mid-IR
continuum luminosities derived in this way are a reliable isotropic
proxy for the intrinsic luminosity of AGN. To do so we have conducted
a UV-to-mid-IR spectral energy distribution decomposition into AGN and
galaxy components. Then we have studied the correspondence between the
AGN luminosities at rest-frame 6\,\mic, corrected for the accretion
disk and host galaxy emission, and at 2-10 keV energies over more than
three orders of magnitude in AGN luminosity. We have used the final
data release of the \swise survey (AllWISE; \citealt{cutri13}) and a
large uniform complete sample of 232 X-ray selected AGN drawn from the
Bright Ultra-hard XMM-Newton Survey. The objects have $z$ in the range
0.05-2.8, and intrinsic (absorption-corrected) 2-10 keV X-ray
luminosities between $10^{42}$ and $10^{46}\lum$. The X-ray
luminosities, corrected for absorption, were computed from a detailed
analysis of the \xmm X-ray spectroscopic data available for all
objects.

This paper is structured as follows. Section\,2 describes the AGN
sample and the multiwavelength data used in this study. In Section\,3
we describe the SED fitting approach used to isolate the AGN mid-IR
emission, the adopted technique to compute rest-frame 6\,\mic\,
luminosities and discuss the effects associated with contamination
from the AGN host galaxies. In Section\,4 we present the relationship
between 6\,\mic\, and 2-10 keV continuum luminosities. The main results
are discussed in Section\,5 and summarized in Section\,6. Throughout
this paper errors are 68 per cent confidence for a single parameter,
and we assume $\Omega_M=0.3$, $\Omega_\Lambda=0.7$ and $H_0={\rm
  70\,km\,s^{-1}\,Mpc^{-1}}$.

\begin{table*}
 \caption{Summary of the main properties of the AGN used in this study.}
 \label{tab0}
 \begin{tabular}{@{}lcccccccc}
  \hline
  Sample & $N$ & $\Delta\,{\rm log}\,L_{\rm 2-10\,keV}$ & $\langle {\rm log}\,L_{\rm 2-10\,keV} \rangle$ & $\Delta\,z$ & $\langle z \rangle$ & $N_{\rm abs}$  \\
   &        & ${\rm erg\,s^{-1}}$ & ${\rm erg\,s^{-1}}$ & & & & \\
   (1)   &  (2)   & (3) &  (4)   &      (5)      &  (6) & (7)\\
  \hline
  All & 232  & 42.12-46.00 & 43.94(43.94)& 0.056-2.860 & 0.558(0.396) & 121(52.1 per cent)\\

  Type\,1 & 137  & 42.32-46.00 & 44.23(44.38)& 0.061-2.860 & 0.713(0.646) & 37(27.0 per cent)\\

  Type\,2 &  95  & 42.12-45.19 & 43.52(43.41)& 0.056-1.266 & 0.334(0.240) & 85(89.5 per cent)\\
  \hline
 \end{tabular}

$Notes$. Column 1: sample; Column 2: number of sources; Column 3: 2-10
 keV X-ray luminosity range (in logarithmic units).  Column 4: Median
 (mean) X-ray luminosity (in logarithmic units); Column 5: redshift range; Column 6: median
 (mean) redshift; Column 7: number (fraction) of objects with detected X-ray absorption.
 \end{table*}

\section[]{AGN sample description}
\label{sample}
Our AGN sample was drawn from the wide-angle Bright Ultra-hard \xmm
Survey (BUXS; \citealt{mateos12}). BUXS is a large, complete,
flux-limited sample of X-ray bright ($f_{\rm 4.5-10 \,keV} >
6\times\,10^{-14}\,\flux$) AGN detected with the \xmm observatory at
4.5-10 keV energies. The survey is based on 381 high Galactic latitude
($\vert$b$\vert$$>$20$^\circ$) \xmm observations having good quality
for serendipitous source detection that were used to derive
extragalactic source count distributions at intermediate X-ray fluxes
(\citealt{mateos08}). BUXS contains 255 AGN, after removal of Galactic
stars and known BL Lacs ($<$3\%), detected over a total sky area of
44.43 deg$^2$.

\subsection[]{UV/optical spectroscopic follow-up and AGN classification}
\label{spectroscopy}
All the sources in BUXS are in the area covered by the Sloan Digital
Sky Survey (SDSS) imaging survey Data Release 7 in the $ugriz$ bands
(\citealt{abazajian09}). To identify the optical counterparts we used
the likelihood-ratio estimator cross-matching algorithm of
\citet{pineau11}\footnote{The code is available from
  http://saada.u-strasbg.fr/docs/fxp/plugin/} (see \citealt{mateos12}
for details). All sources but two have detections in SDSS. For the
remaining two sources we identified their optical counterparts from
our own imaging campaign.

Optical spectroscopic identifications and classifications are
currently available for all but five objects. Such high identification
completeness (98 per cent) guarantees that our study will not suffer
from biases associated with low optical identification rates. The data
has been collected from the literature (mainly from SDSS) and as part
of our extensive optical follow-up identification programme (full
details will be presented in a forthcoming paper).

Throughout this paper, objects with detected UV/optical emission line
velocity widths $\geq$1500 $\kms$ and intermediate Seyfert types 1-1.5
are classified as type 1 AGN (143) while Seyfert types 1.8, 1.9 and 2
and those objects with a galaxy spectrum with no emission lines are
classified as type 2 AGN (107). In this way we have a more uniform
optical spectroscopic classification over the $z$ range sampled by
BUXS. For example, Seyfert 1.9s would be classified as type 2 AGN at
$z$$\gtrsim$0.2, when the H$\alpha$ line at 6563\AA\, is redshifted
from our optical spectra\footnote{The optical spectra from our
  follow-up identification programme do not typically extend to
  wavelengths longer than $\sim$8000\AA.}. We have checked that the
results of our study do not change if we instead classify Seyferts 1.8
and 1.9 as type 1 AGN.

\subsection[]{Broad-band UV-to-mid-IR photometric data}
\label{magnitudes}
As discussed below, one of the main goals of our study is to determine
the rest-frame 6\,\mic\, mid-IR continuum emitted by the AGN dusty torus. To do
so we have decomposed the rest-frame UV-to-mid-IR SEDs of our AGN into
AGN and host galaxy components. To build the SEDs we have collected
all the available photometric data from SDSS, the Two Micron All Sky
Survey (2MASS), the UKIRT Infrared Deep Sky Survey (UKIDSS) and \wise.

We have used SDSS {\tt MODELMAG} magnitudes as they provide reliable
estimates of the fluxes of extended objects while, for point sources,
they are indistinguishable from the standard point spread function
(PSF) magnitudes. We converted SDSS magnitudes and errors into flux
densities following the SDSS
recommendations\footnote{http://www.sdss.org/dr7/algorithms/fluxcal.html}.
A 2 per cent error was added in quadrature to the catalogued flux
errors to account for the uncertainties in the zero-points.

We have used the cross-matching algorithm of \citet{pineau11} to
identify the \swise mid-IR counterparts of our AGN (see
\citealt{mateos12} for details). All sources but six (five type 1 AGN
and one type 2 AGN) have detections with SNR$\geq$2 at the \swise
wavelengths of 3.4, 4.6 and 12\,\mic\, in AllWISE. The mean
X-ray-mid-IR separation is $\lesssim$2 arcsec. Only 10 objects have a
non-zero AllWISE photometric quality flag in at least one of the
\swise bands indicating that the fluxes may be contaminated due to
proximity to an image artefact ({\it $cc\_flags$};
\citealt{cutri13}). After a careful visual check of the \swise images
and SEDs we found no issues in the \swise fluxes for any of the
sources so we decided to keep them in our sample.

We computed flux densities in the \swise bands using either elliptical
aperture magnitudes for the objects flagged as extended (six in total)
or profile fitting photometry for the sources not spatially
resolved. We used the magnitude zero points on the Vega system $F_\nu$
(iso)=309.124 Jy, 171.641 Jy, 30.988 Jy and 8.346 Jy for 3.4, 4.6, 12
and 22\,\mic, respectively (\citealt{wright10}). They correspond to a
spectrum $f_\nu\propto\nu^{\alpha}$ with $\alpha$=-1. We added 1.5 per
cent uncertainties to the catalogued flux errors in all bands to
account for the overall systematic uncertainty from the Vega spectrum
in the flux zeropoints. To account for discrepancies between the red
and blue calibrators used for the conversion from magnitudes to
Janskys we added an additional 10 per cent uncertainty to the 12 and
22\,\mic\, catalogued fluxes \citep{wright10}.

At the stage of the project when we searched for the near-IR
photometric information available for all the AGN in BUXS we had
already accurate source positions for the optical counterparts of all
the objects in the sample. Therefore, to identify the near-IR
counterparts of our AGN, we cross-correlated the coordinates of their
optical counterparts with the Two Micron All Sky Survey (2MASS)
catalogues of point and extended sources (\citealt{jarrett00};
\citealt{cutri03}) and the UKIRT Infrared Deep Sky Survey (UKIDSS)
Data Release 10 catalogues (\citealt{lawrence07}) using a simple
position matching with a search radius of 2 arcsec. In total 166
objects ($\sim$71 per cent) have detections in the near-IR. The mean
optical-infrared separation is $<$1 arcsec. We have checked that
increasing the search radius does not produce any additional
matches. From UKIDSS we used Petrosian magnitudes for extended sources
and the default aperture magnitudes for point sources. As for SDSS, we
added a 2 per cent error in quadrature to the catalogued flux errors
to account for uncertainties in the zero-points. 2MASS and UKIDSS
magnitudes were converted into fluxes using the zero points
$F_\nu$(iso)=1594 Jy, 1024 Jy and 666.7 Jy for the 2MASS $JHK_s$
(\citealt{cohen03}) and $F_\nu$(iso)=2232 Jy, 2026 Jy, 1530 Jy, 1019
Jy and 631 Jy for the UKIDSS $ZYJHK$ (\citealt{hewett06}).

We note that if we use the \swise magnitude zero points in the Vega
system corresponding to the spectral shapes assumed to convert
magnitudes to flux densities in UKIDSS ($\sim\nu^{0}$) and 2MASS
($\sim\nu^{-2}$), the differences in the derived flux densities in the
\swise bands would be lower than one per cent at 3.4, 4.6 and
22\,\mic\, and nine per cent at 12\,\mic, all below the systematic
uncertainties in the catalogued fluxes.

\subsection[]{X-ray spectral properties}
\label{xspec}
Here we give a brief summary of the analysis conducted to determine
the X-ray properties of our AGN. Full details on the X-ray spectral
extraction procedure, analysis of the data and X-ray properties
of the full sample of BUXS AGN will be presented in a forthcoming
paper. 

We have good quality\xmm X-ray spectra in the observed energy range
from 0.25 to 10 keV for all the AGN in this study. The median (mean)
background subtracted number of counts in the 0.25-10 keV EPIC spectra
(MOS+pn) is 1443 (3127), where the counts range from 52 to 75,000. The
X-ray spectroscopic analysis was conducted with the Xspec package
(v12; \citealt{arnaud96}). For each object we fitted the data with a
combination of different models to determine the shape of the direct
and scattered continuum components (modelled with power-laws) and
rest-frame line-of-sight X-ray absorption (considering both full
coverage and partial coverage of the continuum source). All
  models include the Galactic absorption for each source using column
  densities taken from \citet{dickey90}. We have also accounted for
the 'soft excess' i.e., any emission at energies $\lesssim$ 1 keV in
excess of the extrapolation of the primary continuum, an almost
ubiquitous feature in the X-ray spectra of AGN
(e.g. \citealt{turner97}; \citealt{porquet04}; \citealt{guainazzi05};
\citealt{piconcelli05}; \citealt{mateos10}; \citealt{scott12};
\citealt{winter12}) using thermal and black-body models. To accept the
detection of a model component we used the F-test with a significance
threshold of 95 per cent.

\begin{figure}
  \centering
  \begin{tabular}{cc}
    \hspace{-0.7cm}\includegraphics[angle=90,width=0.496\textwidth]{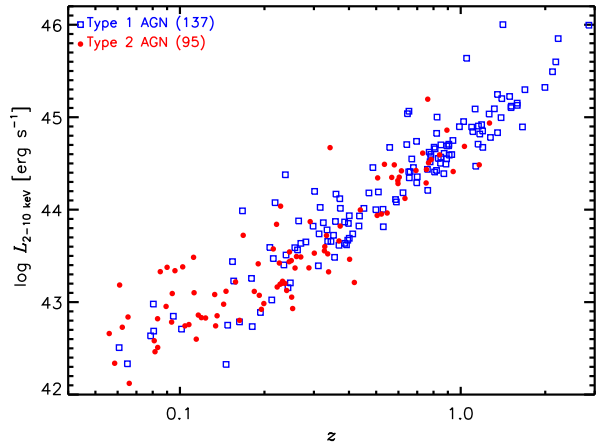}
  \end{tabular}
  \caption{Hard X-ray luminosity versus $z$ for the AGN in BUXS
    selected for this study.}
  \label{fig1}
\end{figure}

If X-ray absorption was not detected we computed 1$\sigma$ upper
limits using the best-fitting model parameters. Though BUXS includes
many objects with X-ray absorption close to being Compton-thick, when
taking into account the uncertainties in the best-fit rest-frame
column densities, all X-ray absorption values are consistent with
being in the Compton-thin regime (${\rm N_H<1.5\times
  10^{24}\,cm^{-2}}$). Thus, based on the existing X-ray data we do
not have any securely confirmed Compton-thick AGN. The X-ray
luminosities used in this paper have been determined from the best-fit
models in the rest-frame 2-10 keV energy band and are corrected for
X-ray absorption. Examples of our X-ray spectral modelling are shown
in Fig.~\ref{fig2}.

\subsection[]{AGN sample selection}
\label{agn_sample}
For the study presented here we have selected the 232 AGN in BUXS with
absorption-corrected X-ray luminosities $\geq$$10^{42}\lum$ and \swise
detections in the three shortest wavelength bands (3.4, 4.6 and
12\,\mic) with SNR$\geq$2. Out of these, 137 objects are type 1 AGN
(122 Seyfert 1s, 15 Seyfert 1.5s) and 95 type 2 AGN (3 Seyfert 1.8s,
14 Seyfert 1.9s and 78 Seyfert 2s). The X-ray luminosity limit has
been imposed to reduce to a minimum the effects of host galaxy
contamination in the mid-IR regime by increasing the contrast of the
AGN contribution.

The selected AGN have 2-10 keV X-ray luminosities between $10^{42}$
and $10^{46}{\rm \,erg\,s^{-1}}$ and $z$ in the range
0.05-2.8. Table~\ref{tab_Appendix} lists the optical spectroscopic
classification and intrinsic X-ray luminosity for all the AGN used in
our study while Table~\ref{tab0} summarizes their main properties. The
distribution of their $z$ and 2-10 keV luminosities is shown in
Fig.~\ref{fig1}.

\section{AGN mid-IR continuum luminosities}
\label{l6}
Our study aims to investigate how well we can isolate the mid-IR
emission associated with the AGN dusty torus using \swise broad band
photometric data, and whether the mid-IR AGN luminosities determined
in this way are a reliable isotropic proxy of the AGN intrinsic
power. To do so we have studied the correspondence between the
rest-frame 6\,\mic\, and 2-10 keV X-ray continuum luminosities for our
complete, flux-limited sample of X-ray selected AGN.

We have used rest-frame 6\,\mic\, monochromatic luminosities as a
tracer of the AGN dusty torus emission. As such emission can be well
approximated by a power-law (see, e.g. \citealt{neugebauer79};
\citealt{elvis94}; \citealt{richards06}; \citealt{assef10}), to
compute the 6\,\mic\, luminosities we have used a simple linear
interpolation/extrapolation in log-log space between the \swise
catalogued fluxes at 4.6 and 12\,\mic. Sec.~\ref{decomp} describes in
detail how we corrected these luminosities for both host galaxy
contamination and for the emission from the AGN accretion disk.

We have not used the \swise 3.4\,\mic\, fluxes because, as shown in
Sec.~\ref{decomp}, they are severely contaminated by
starlight. Moreover, we have not used the \swise 22\,\mic\, fluxes
because, as this survey is much shallower than the ones conducted at
shorter wavelengths, many of our objects are not detected. Indeed, out
of the 232 AGN selected for this study just 168 (72.4 per cent) are
detected with SNR$\geq$2 at 22\,\mic.

\begin{figure*}
  \centering
  \begin{tabular}{cc}
    \hspace{-0.7cm}\includegraphics[angle=-90,width=0.46\textwidth]{fig2a.ps}
    \hspace{0.7cm}\includegraphics[angle=-90,width=0.46\textwidth]{fig2b.ps}\\
    \hspace{-0.7cm}\includegraphics[angle=90,width=0.48\textwidth]{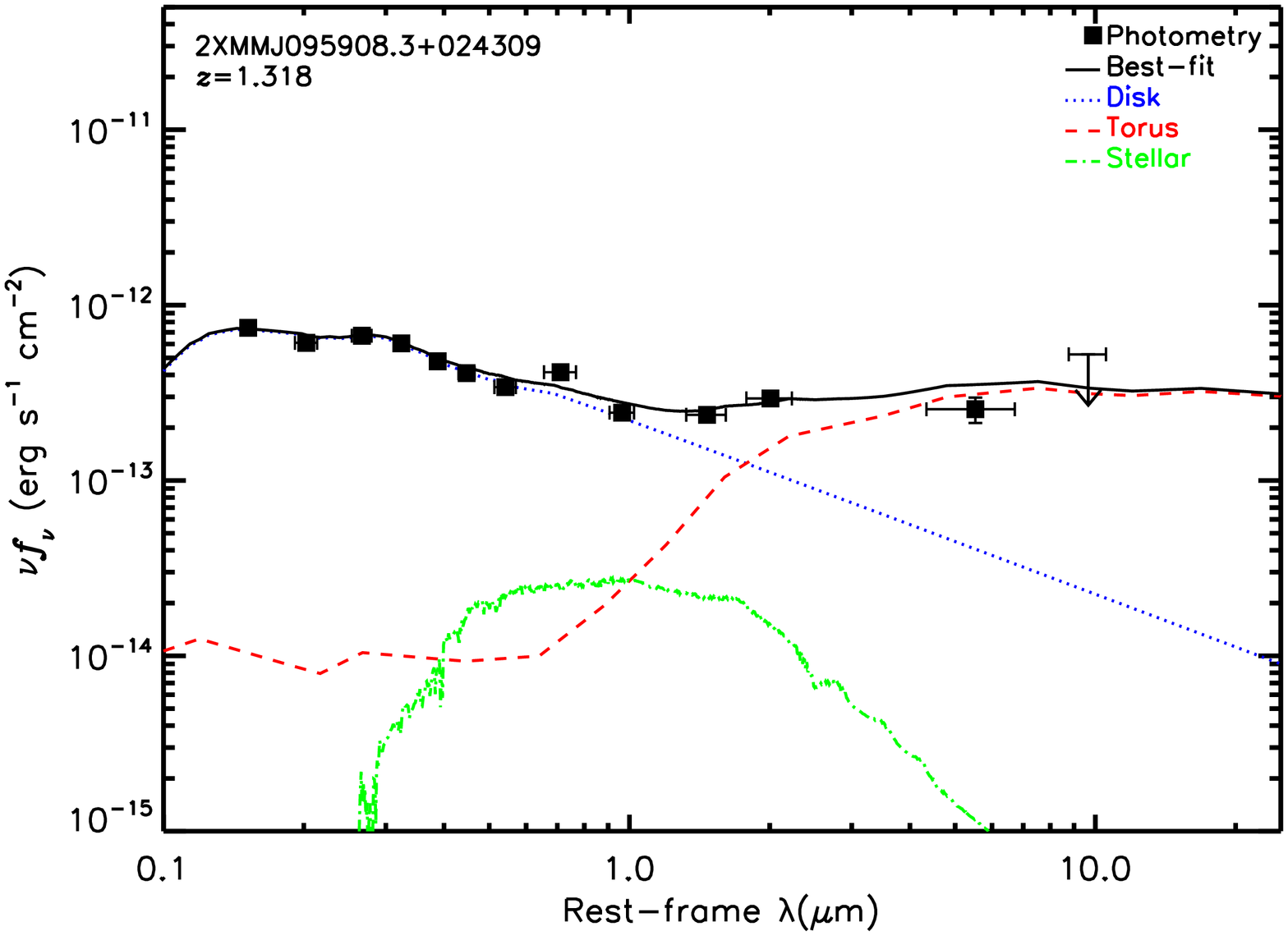}
    \hspace{0.7cm}\includegraphics[angle=90,width=0.48\textwidth]{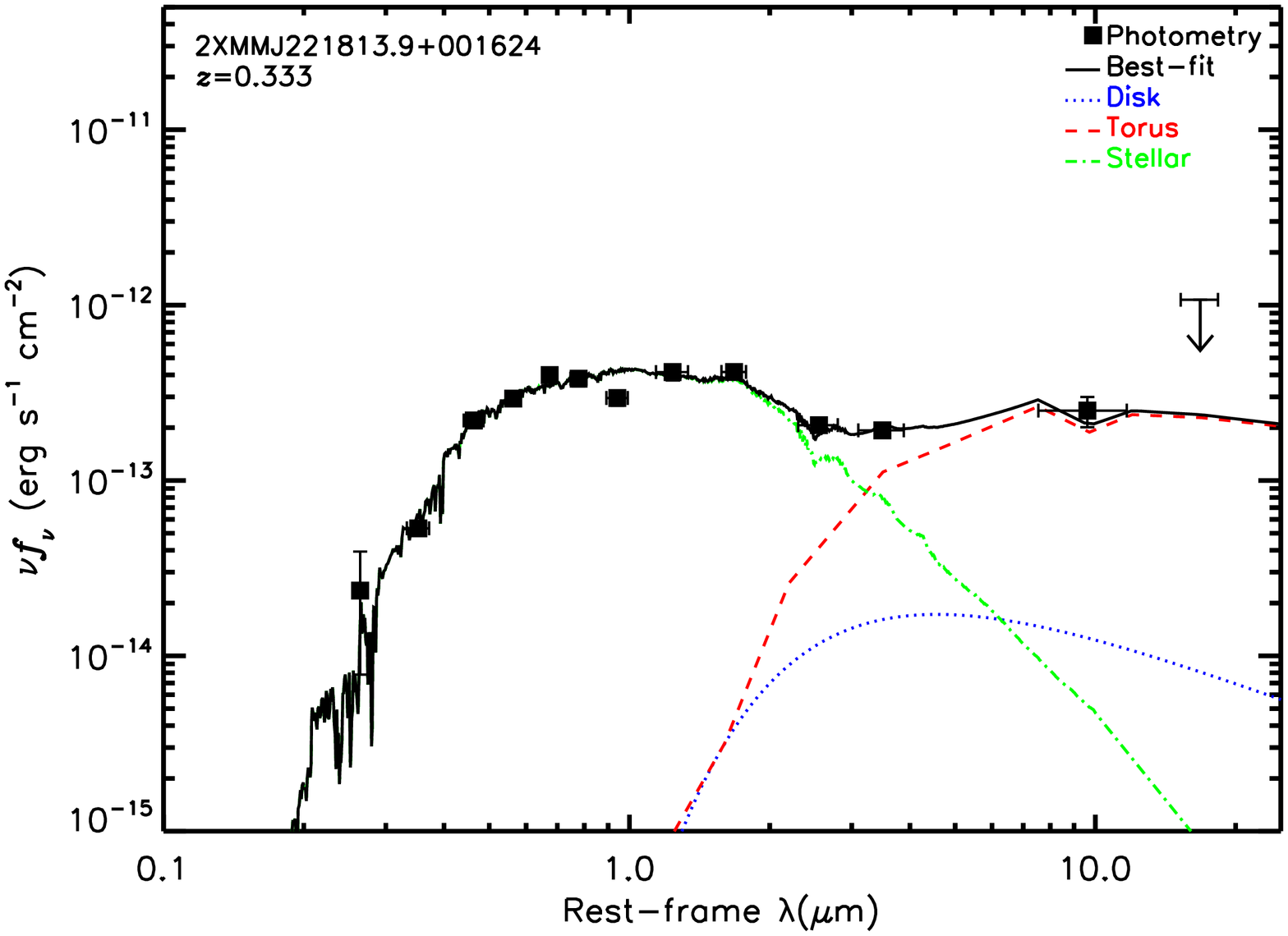}\\
  \end{tabular}
  \caption{Examples of the modelling of the \xmm EPIC X-ray
      spectroscopic data and SED decomposition for AGN classified as
      type 1 (left plots) and type 2 (right plots), respectively. Top:
      The X-ray spectra for 2XMMJ095908.3+024309 are best modelled
      with a simple unabsorbed power-law model (broad-band photon
      index $\Gamma$=1.88$\pm$0.02; solid lines). For
      2XMMJ221813.9+001624 the best-fit model consists on an absorbed
      power-law ($\Gamma$=1.77$_{-0.37}^{+0.39}$ and ${\rm N_H}$=${\rm
        40.4_{-7.1}^{+8.0}}$, where the column density is in units of
      ${\rm 10^{22}\,cm^{-2}}$) plus a low-energy scattered power-law
      with same spectral index as that of the direct absorbed
      component (dotted lines). Bottom: The figures clearly illustrate
      that, at the shortest wavelength bands of \wise, the emission
      from the AGN host galaxies can dominate over that of the dusty
      torus especially in type 2 AGN where the extinction of the AGN
      emission is systematically higher.}
  \label{fig2}
\end{figure*}

\subsection{SED decomposition into the AGN and their hosts}
\label{decomp}
Even at the high AGN luminosities involved in our study, the 4.6 and
12\,\mic\, fluxes (rest-frame $\sim$1.5-4\,\mic\, at $z$$\sim$2) might
be contaminated by the emission from the AGN host galaxies. To isolate
the AGN component and derive the 6\,\mic\, luminosity associated with
the dusty torus, we have decomposed the rest-frame UV-to-mid-IR SEDs of
our objects into AGN and galaxy emission. We used the best-fit results
to correct the 6\,\mic\, luminosities for contamination from the AGN
host galaxies and accretion disk emission. We stress that the main
goal of the SED fitting analysis is to properly model all the
continuum components in the observed SEDs to ensure a reliable
determination of the dusty torus emission. We have used the fitting
code SEd Analysis using BAyesian Statistics
(SEABASs)\footnote{http://astro.dur.ac.uk/$\sim$erovilos/SEABASs/}
described in detail in \citet{rovilos14}. The program uses the maximum
likelihood method and a Monte Carlo Markov chain (MCMC) sampling
technique to determine both the optimum combination of SED templates
and the associated uncertainties.

\subsubsection{SED components}
\label{components}
We fitted the SEDs with three model components to reproduce the
emission from the AGN accretion disk, the dusty torus and the stellar
emission of the hosts. 

To describe the UV/optical emission from the accretion disk we have
used the average type 1 quasar SED of \citet{richards06} at
$\lambda$$<$0.7\,\mic\, and a power-law $\lambda f_\lambda \propto
\lambda^{-1}$ at longer wavelengths. To redden the accretion disk we
have used the \citet{gordon98} Small Magellanic Cloud (SMC) extinction
law at $\lambda$$<$3300 \AA, as it best represents the nuclear
reddening of quasars (\citealt{hopkins04}), while at $\lambda$$>$3300
\AA\, we used the Galactic extinction law from \citet{cardelli89}. In
both cases we assume $R_V$=3.1.

We have characterized the mid-IR emission from the AGN dusty torus
with the template library of \citet{silva04}. The library includes
four templates divided according to the amount of X-ray
absorption. These templates were generated by combining the nuclear
mid-IR observations of a large sample of Seyfert galaxies and
interpolation using the smooth-density dust emission model of
\citet{granato94}. Of these we have used the Seyfert 1 template and
the two Seyfert 2 templates corresponding to X-ray column densities in
the Compton-thin regime: ${\rm 10^{22}<N_{H}<10^{23}\,cm^{-2}}$ and
${\rm 10^{23}<N_{H}<10^{24}\,cm^{-2}}$. We have not used either the
optical classification or best-fit X-ray column densities to choose
amongst torus templates i.e., we have considered all three torus
templates to fit the SEDs of our type 1 and type 2 AGN.

Finally, we have reproduced the emission from the stellar population
of the AGN hosts at rest-frame optical-near-IR wavelengths with the
stellar population synthesis models of \citet{bruzual03}. We used a
library of 75 stellar templates with solar metallicity and a Chabrier
initial mass function (\citealt{chabrier03}). We have used ten
exponentially decaying star formation histories with characteristic
times $\tau$=0.1 to 30 Gyr and a model with constant star formation,
and a set of ages in the range 0.1 to 13 Gyr (see \citealt{lusso13}
for a similar approach). We have checked that none of our fits require
such long ($>10$ Gyr) and probably unphysical ages. We reddened the
templates using the \citet{calzetti00} dust extinction law and a range
of E(B-V)$_{\rm gal}$ from 0 to 2.

To avoid degeneracies between the AGN and galaxy templates, we have
ignored any possible contamination from dust heated by star formation
in the AGN hosts. As demonstrated in Sec.~\ref{host_cont_mir}, at the
high AGN luminosities involved in our study and rest-frame wavelengths
sampled with the \swise 4.6 and 12\,\mic\, surveys, contamination due
to star formation has a negligible effect on our
results. Nevertheless, as the star formation emission rapidly
increases at rest-frame wavelengths $\gtrsim$8\,\mic, contamination of
the \swise 22\,\mic\, fluxes might not be negligible at the lowest
redshifts (and AGN luminosities). To minimize the impact of such
effect we have treated all 22\,\mic\, detections (168 in total) as
upper limits.

\begin{table*}
 \caption{Contribution from the reddened accretion disk and the AGN
   hosts to the mid-IR at rest-frame 6\,\mic.}
 \label{tab01}
 \begin{tabular}{@{}lcccccccc}
  \hline
  Sample & $N$ & $\Delta\,{\rm log\,}L_{\rm 2-10\,keV}$ & $f_{6\,\mu {\rm m}}^{\rm st}$ & $f_{6\,\mu {\rm m}}^{\rm disk}$ & $f_{6\,\mu {\rm m}}^{\rm disk+st}$ & $f_{6\,\mu {\rm m}}^{\rm sb}$\\
   &        & ${\rm erg\,s^{-1}}$ & per cent & per cent  & per cent & per cent \\
   (1)   &  (2)   & (3) &  (4)   &      (5) &     (6)  &   (7) \\
  \hline
  All    & 36  & 42-43 & $20.76_{-15.88}^{+22.81}$ & $3.66_{-2.47}^{+11.57}$ & $28.13_{-11.23}^{+17.94}$ & $7.93_{-4.79}^{+8.92}$ \\
  Type 1 & 12  & 42-43 & $10.95_{-8.13}^{+16.97}$ & $15.23_{-5.72}^{+14.96}$ & $27.15_{-10.25}^{+18.46}$ & $7.52_{-5.38}^{+9.32}$ \\
  Type 2 & 24  & 42-43 & $26.02_{-14.94}^{+19.95}$ & $2.94_{-2.16}^{+2.32}$  & $29.25_{-12.37}^{+17.13}$ & $9.53_{-5.60}^{+7.72}$ \\ \\

  All    & 86  & 43-44 & $7.81_{-5.10}^{+8.93}$ & $7.39_{-4.07}^{+7.18}$     & $17.96_{-6.36}^{+8.73}$ & $2.54_{-1.89}^{+3.20}$ \\
  Type 1 & 39  & 43-44 & $5.09_{-3.99}^{+6.78}$ & $10.79_{-5.23}^{+6.09}$    & $18.32_{-8.54}^{+8.55}$  & $1.65_{-1.18}^{+2.56}$ \\ 
  Type 2 & 47  & 43-44 & $11.57_{-6.12}^{+5.91}$ & $5.94_{-3.18}^{+3.83}$    & $17.41_{-5.63}^{+8.26}$ & $3.19_{-1.91}^{+3.22}$ \\ \\

  All    & 88  & 44-45 & $2.12_{-2.12}^{+2.30}$ & $10.82_{-5.87}^{+5.25}$    & $13.84_{-4.34}^{+7.08}$ & $0.37_{-0.25}^{+0.93}$ ($3.72_{-2.46}^{+9.29}$)\\ 
  Type 1 & 65  & 44-45 & $1.40_{-1.40}^{+2.67}$ & $11.40_{-5.22}^{+6.74}$    & $14.18_{-4.68}^{+6.88}$ & $0.32_{-0.21}^{+0.62}$ ($3.25_{-2.14}^{+6.17}$)\\
  Type 2 & 23  & 44-45 & $4.40_{-2.88}^{+5.18}$ & $8.73_{-4.11}^{+3.60}$     & $13.17_{-2.91}^{+4.10}$ & $0.63_{-0.44}^{+1.33}$ ($6.27_{-4.45}^{+13.30}$)\\ \\

  Type 1 & 21  & 45-46 & $0.23_{-0.23}^{+0.83}$ & $11.33_{-5.54}^{+4.23}$    & $12.45_{-6.20}^{+6.94}$ & $0.01_{-0.01}^{+0.05}$ ($0.11_{-0.06}^{+0.65}$)\\
  \hline
 \end{tabular}

$Notes$. Column 1: sample; Column 2: number of sources; Column 3: 2-10
 keV X-ray luminosity range (in logarithmic units).  Columns 4 and 5:
 median contribution from the AGN hosts stellar emission and the
 reddened accretion disk at rest-frame 6\,\mic, respectively. Column
 6: median of the sum of the contributions from the AGN hosts and the
 reddened accretion disk at rest-frame 6\,\mic. Column 7: (very
 conservative) median upper limits for the level of contamination from
 star formation at rest-frame 6\,\mic\, assuming that the AGN hosts
 are LIRGs or ULIRGs (values in brackets; see Sec.~\ref{host_cont_mir}
 for details). The error bars represent the 16 and 84 percentiles. In
 the cases where the lower limit of the error bar reached zero, the
 upper limit corresponds to the 68 percentile (i.e., smallest interval
 that encloses the 68 per cent probability). Please note that to
 compute the numbers we have not included the only type 2 AGN in the
 sample with $L_{\rm 2-10\,keV}>10^{45}\lum$.
 \end{table*}

In the following subsections, we discuss in more detail the SED
fitting procedure and summarize the main results. In the AGN with
signatures of the Broad Line Region in their rest-frame UV/optical
spectrum (i.e. Seyfert types 1 to 1.9) the contamination from the
accretion disk at UV/optical and near-IR wavelengths is not
negligible. Therefore we have carried out the SED fits following a
different approach for AGN with detected UV/optical broad emission
lines (i.e. where we have a relatively unobscured view of the BLR and
accretion disk; Sec.~\ref{sed_bl}) and for AGN without detected
UV/optical broad emission lines (i.e. where the accretion disk
emission is highly obscured; Sec.~\ref{sed_nl}). Nevertheless, we
remind the reader that, to have a more uniform optical spectroscopic
classification over the $z$ range sampled by BUXS, throughout the
paper we have classified Seyfert 1.8-1.9 (17 in total) as type 2 AGN
(see Sec.~\ref{spectroscopy}).

\subsubsection{SED fitting of AGN with detected UV/optical broad emission lines} 
\label{sed_bl}
For the type 1 AGN (137) and intermediate Seyfert types 1.8-1.9 (17)
we have used a range of reddening values for the accretion disk,
E(B-V)$_{\rm disk}$, between 0 and 0.65 with a step
$\Delta$E(B-V)$_{\rm disk}$=0.01. As indicated in
\citet{caccianiga08}, the separation between optical type 1 and type 2
AGN corresponds to $A_{\rm V}$$\sim$2 mag, or E(B-V)$\sim$0.65
assuming a Galactic standard conversion. To take into account any
intrinsic dispersion in the SEDs of our AGN, the normalizations of the
disk and torus components were left free to vary. The SED fitting was
conducted both with and without the stellar component. We accepted the
detection of the stellar component if the $\chi^2$ difference between
models, $\Delta\chi^2$, was higher than 6.17 (2$\sigma$ confidence for
two parameters)\footnote{The relationship between $\Delta\chi^2$ and
  the log-likelihood difference is $\Delta\chi^2=-2\Delta({\rm ln} L)$
  (see Appendix A in \citealt{rovilos14} for details).}. Otherwise we did
not apply any correction for host galaxy contamination to the observed
\swise fluxes.
 
We have found a median (mean) nuclear extinction E(B-V)$_{\rm
  disk}$=0.05 (0.11), with 68 per cent of the sources having
E(B-V)$_{\rm disk}$ below 0.1. Considering only those sources showing
evidence for both X-ray absorption (see Sec.~\ref{xspec}) and
UV/optical reddening (49 in total), we found that the great majority
(88 per cent) have dust-to-gas ratios E(B-V)/N$_{\rm H}$ lower than
Galactic by a factor varying from 1.5 to more than 100 in a few cases,
in good agreement with \citet{maiolino01}. In 42 objects (38 with
$L_{\rm 2-10\,keV}>10^{44}\lum$) adding the stellar component did not
improve the quality of the fit above the selected threshold. The
median (mean) contribution of the accretion disk at rest-frame 6\,\mic,
relative to the best-fit torus model, is $11.8_{-6.1}^{+7.6}$
(12.9$\pm$7.8) per cent and has no significant luminosity dependence
based on the Spearman's rank test (correlation coefficient
$\rho$=-0.12 and probability to reject the null hypothesis 86.7 per
cent). The median uncertainty associated with the correction for the
accretion disk emission determined from the SED fits is 13 per
cent. Such uncertainties were added in quadrature to our 6\,\mic\,
luminosity estimates.

\begin{figure}
  \centering
  \begin{tabular}{cc}
    \hspace{-0.7cm}\includegraphics[angle=90,width=0.496\textwidth]{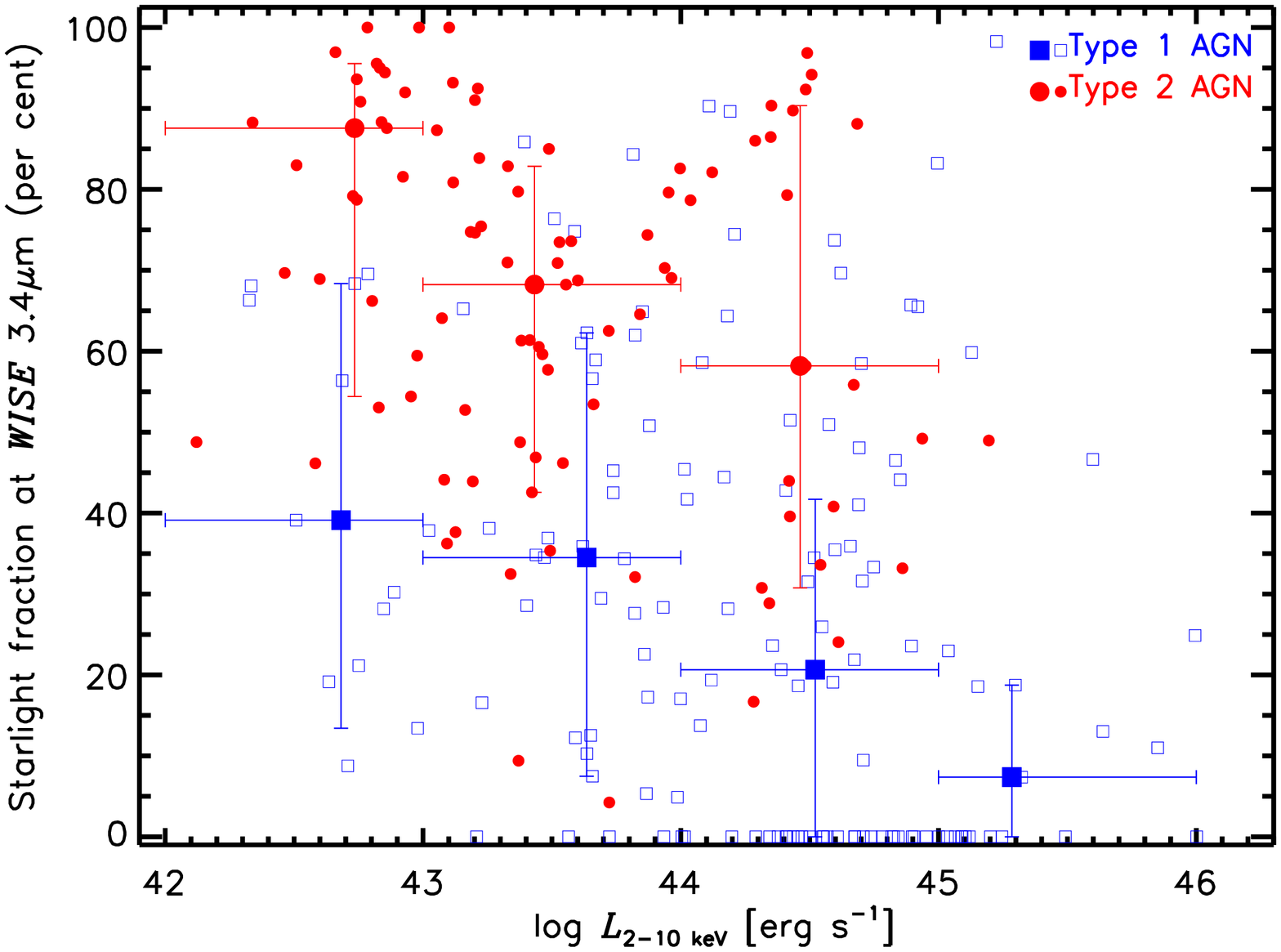}\\
    \hspace{-0.7cm}\includegraphics[angle=90,width=0.496\textwidth]{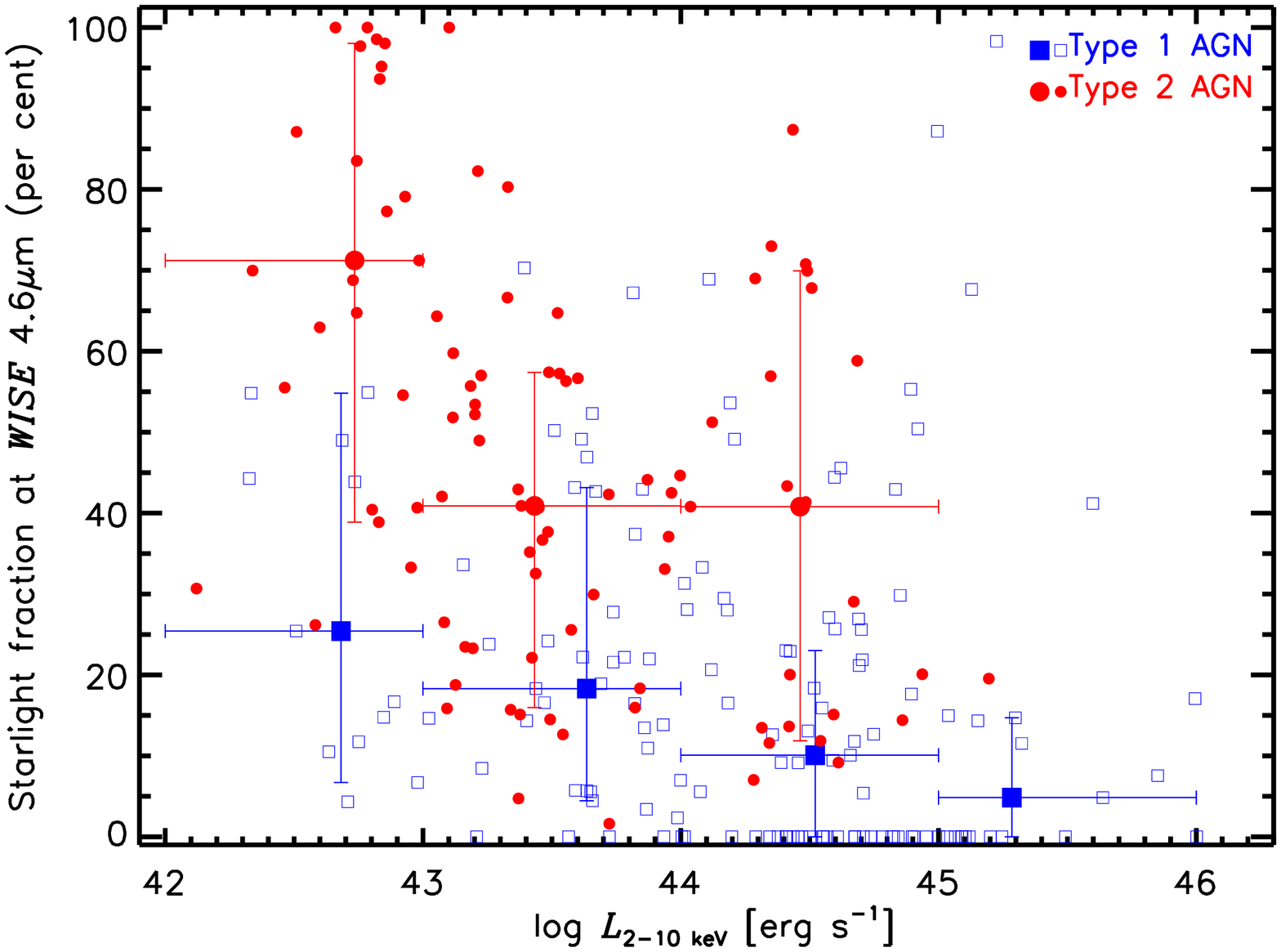}\\
    \hspace{-0.7cm}\includegraphics[angle=90,width=0.496\textwidth]{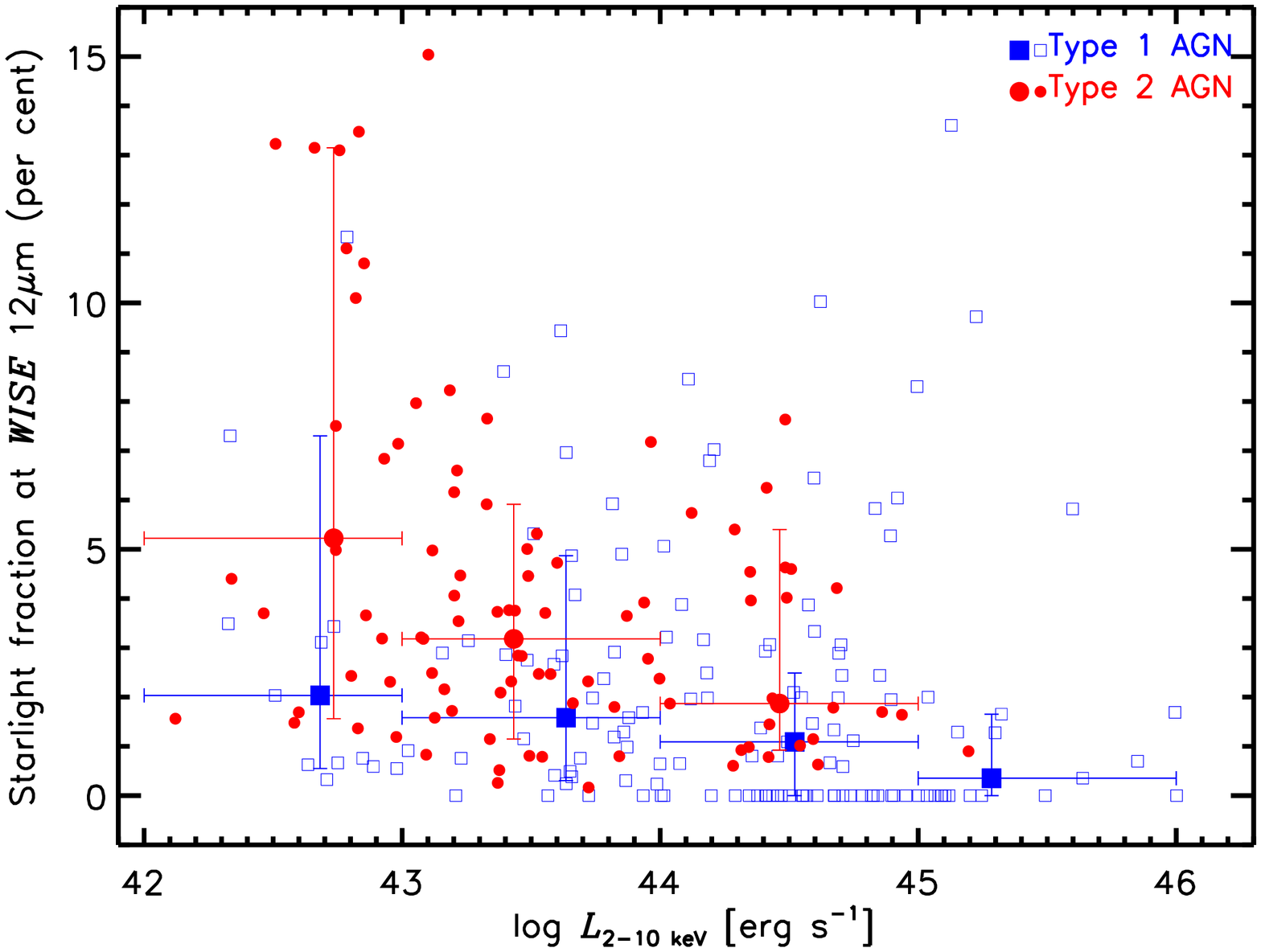}
  \end{tabular}
  \caption{Contribution from the AGN host galaxies to the mid-infrared
    fluxes measured with \swise at 3.4, 4.6 and 12\,\mic\, (observed
    frame). Large filled symbols and vertical error bars indicate the
    median and 16 and 84 percentiles (68 per cent enclosed, equivalent
    to 1$\sigma$) of the distribution of values in different
    luminosity intervals indicated with the horizontal error bars. In
    the cases where the lower limit of the error bar reached zero, the
    upper limit corresponds to the 68 percentile (i.e., smallest
    interval that encloses the 68 per cent probability).}
  \label{fig3}
\end{figure}

\subsubsection{SED fitting of AGN without detected UV/optical broad emission lines} 
\label{sed_nl}
In the 78 AGN without detected broad emission lines in their
rest-frame UV/optical spectra the accretion disk emission is highly
obscured and the observed fluxes at rest-frame UV-to-near-IR
wavelengths are dominated by the AGN host galaxies. Nevertheless, for
AGN with nuclear reddening levels A$_{\rm V}$$\sim$2-5 mag, the accretion
disk emission at near-IR wavelengths ($\sim$1-2\,\mic) could still be
relevant. As for these objects we do not have a direct view of the
accretion disk, we cannot constrain both the extinction and
normalization of such a component. The approach we followed was to fix
the normalization of the accretion disk relative to the torus
component to the median value obtained for the AGN with detected
UV/optical broad lines. We then created a library of reddened
accretion disk templates using a range of E(B-V)$_{\rm disk}$ values
between 0.65 and 10 with a step $\Delta$E(B-V)$_{\rm
  disk}$=0.15. These templates were added to each of the three torus
templates used for the SED fitting. We repeated the fitting process,
fixing the accretion disk normalization relative to the torus
component to the median-1$\sigma$ and median+1$\sigma$ values, and
used the best-fit results to determine the uncertainties in the
correction of the 6\,\mic\, luminosities for the accretion disk
emission.

The best-fit E(B-V)$_{\rm disk}$ values have a nearly uniform
distribution between the chosen limits with a median (mean) value of
E(B-V)$_{\rm disk}$=6.22 (5.95). We stress, however, that the measured
nuclear extinctions should be regarded as less accurate than for
objects where we have a direct view of the accretion disk. The median
(mean) contribution of the reddened accretion disk at rest-frame
6\,\mic, relative to the best-fit torus model, is $6.5_{-2.6}^{+3.9}$
(6.9$\pm$3.3) per cent. The median uncertainty associated with the
correction for the accretion disk emission from the SED fits is 46 per
cent. Although as pointed out previously, the extinction values are
less accurate, we found that 67 per cent of the sources have
dust-to-gas ratios lower than Galactic by a factor that varies from
$\sim$1.5 to 90, again in very good agreement with \citet{maiolino01}.

\begin{figure}
  \centering
  \begin{tabular}{cc}
    \hspace{-0.7cm}\includegraphics[angle=90,width=0.496\textwidth]{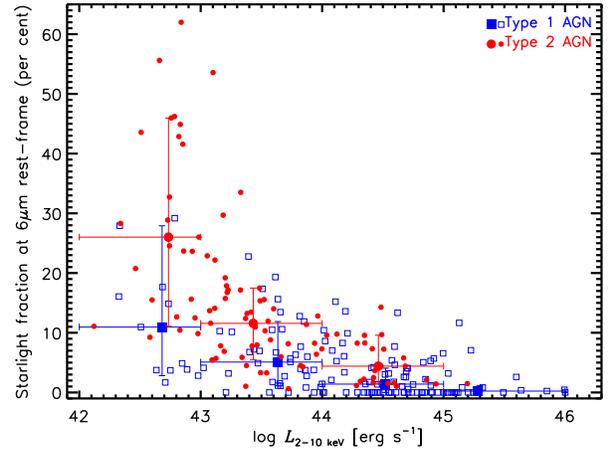}
  \end{tabular}
  \caption{Host galaxy contribution to the rest-frame 6\,\mic\,
    luminosities as a function of the AGN luminosity. Large filled
    symbols and vertical error bars indicate the median and the 16 and
    84 percentiles (68 per cent enclosed, equivalent to 1$\sigma$) of
    the distribution of values in different luminosity intervals
    indicated with the horizontal error bars. In the cases where the
    lower limit of the error bar reached zero, the upper limit
    corresponds to the 68 percentile (i.e., smallest interval that
    encloses the 68 per cent probability).}
  \label{fig4}
\end{figure}

Examples of the SED fitting procedure for type 1 and type 2 AGN are
shown in Fig.~\ref{fig2}. The contamination from the AGN host galaxies
at rest-frame near-IR and mid-IR wavelengths is discussed in
Sec.~\ref{host_cont_nir} and Sec.~\ref{host_cont_mir} and summarized
in Fig.~\ref{fig3} and Fig.~\ref{fig4}. In Table~\ref{tab01} we report
the median contribution from the AGN hosts and the reddened accretion
disk at rest-frame 6\,\mic\, for type 1 and type 2 AGN at different
X-ray luminosities.

\subsection{Stellar contamination in the rest-frame near-IR and mid-IR}
\label{host_cont_nir}
 We have corrected our rest-frame 6\,\mic\, luminosities for the
 Rayleigh-Jeans tail of the emission of the different underlying
 stellar populations of the AGN host galaxies as follows. First, we
 have computed the observed fluxes in the 4.6 and 12\,\mic\, passbands
 associated with starlight emission using the best-fit stellar
 templates. To accurately reproduce the measurement process with the
 wide bandpasses of the \swise filters we have convolved the best-fit
 stellar SEDs with the \swise relative spectral response (RSR) filter
 curves. Then we have determined the luminosity at 6\,\mic\, associated
 with starlight emission through linear interpolation/extrapolation in
 log-log space. The median uncertainty associated with the correction
 for stellar emission determined from the SED fits is 19 per
 cent. These were added in quadrature to our 6\,\mic\, luminosity
 uncertainties.

Fig.~\ref{fig3} shows the relative contribution from starlight to the
fluxes measured with \swise at 3.4, 4.6 and 12\,\mic\, as a function of
AGN luminosity. In the cases where the emission from the AGN hosts was
not detected with sufficient significance ($\Delta\chi^2$$<$6.17) we
assigned a zero value to the fractions. For AGN optically classified
as type 2, the starlight contamination to the \swise fluxes at 3.4 and
4.6\,\mic\, is important across the full luminosity range sampled by
our study. This is expected as dust extinction of the AGN emission is
more important in type 2 objects. In addition, in flux-limited
surveys, such as BUXS, the most luminous objects are detected at
higher $z$ and the \swise passbands are sampling shorter rest-frame
wavelengths where the starlight emission peaks. As dust extinction of
the AGN emission is less severe in type 1 objects, the starlight
contamination to their mid-IR fluxes is significantly smaller than
that for type 2 AGN but still non-negligible, especially at 3.4\,\mic\,
(\citealt{kotilainen92}; \citealt{alonso-herrero96};
\citealt{alonso-herrero03}). Only the \swise fluxes at 12\,\mic\, are
dominated by the AGN at all luminosities, at least up to
$z$$\sim$2. An alternative way to visualize the dependence of the
level of stellar contamination to the \swise fluxes on the AGN
luminosity and optical class is representing the distribution of
\swise colours for our sources. If we adopt the mid-IR colour
definition of \citet{mateos12} (i.e. ${\rm log({\it f}_{\rm 4.6\,\mu
    m}/{\it f}_{\rm 3.4\,\mu m})}$ vs. ${\rm log({\it f}_{\rm 12\,\mu
    m}/{\it f}_{\rm 4.6\,\mu m})}$ flux density ratios), the most
contaminated objects should be located in the bottom-right part of the
colour-colour diagram. Indeed as we see in Fig. 6 in \citet{mateos12}
the less luminous objects, in particular those classified as type 2
AGN, have overall the bluest ${\rm log({\it f}_{\rm 4.6\,\mu m}/{\it
    f}_{\rm 3.4\,\mu m})}$ mid-IR colours.

Fig.~\ref{fig4} illustrates the correction for starlight emission to
our rest-frame 6\,\mic\, luminosities computed as indicated above. The
corresponding values are reported in Table~\ref{tab01}.

\subsection{Contamination from star formation in the rest-frame mid-IR}
\label{host_cont_mir}
As indicated in Sec.~\ref{components}, we have not corrected our
6\,\mic\, luminosities for the emission from dust in star forming
regions in the AGN hosts. For example, Polycyclic Aromatic Hydrocarbon
(PAH) features can probe star formation in the AGN hosts over time
scales of tens of millions of years (e.g. \citealt{peeters04}).
Nevertheless, it is expected that any PAH redshifted into the \swise
observed bands superimposed on the strong AGN continuum would be
smoothed out due to the wide bandpasses of the filters, especially for
the extremely wide 12\,\mic\, filter. To demonstrate that this is the
case, we have determined upper limits on the level of contamination
from star formation to our 6\,\mic\, luminosities. To do so we have
used the main sequence and starburst galaxy templates from
\citet{elbaz11}. For each object in our sample we have determined the
upper limit in the normalization of the starforming templates assuming
that the hosts of our AGN are Luminous Infrared Galaxies (LIRGs) with
$L_{\rm 8-1000\,\mu m}$$\leq$${\rm 6\times 10^{11}\,}L_\odot$ and star
formation rates (SFRs) of $\sim$100 $M_\odot\,{\rm yr^{-1}}$ based on
the \citet{kennicutt98} relation. The less than equal sign indicates
the additional constraint that none of the fluxes determined from each
template in the \swise bands (taking into account the filter RSRs)
should be above the observed values. Then we computed the 6\,\mic\,
luminosities from linear interpolation/extrapolation of the 4.6 and
12\,\mic\, fluxes in log-log space as we did for the original
data. Following this approach we found that the contribution from star
formation at 6\,\mic\, is well below 15 per cent for 83.3 per cent of
the AGN with $L_{\rm 2-10\,keV}$ in the range $10^{42}$-$10^{43}{\rm
  \,erg\,s^{-1}}$ and for $>$99 per cent of the sources with $L_{\rm
  2-10\,keV}$$\geq$$10^{43}{\rm \,erg\,s^{-1}}$.

\begin{table*}
 \caption{Summary of the results of the linear regression analysis.}
 \label{tab1}
 \begin{tabular}{@{}lcccccccccc}
  \hline
  Sample & Method & $N$ & $\rho$ ($p$-value) & $\tau$ ($p$-value) & $\sigma_{\rm obs}$ & $\sigma_{\rm int}$ & $\alpha$ & $\beta$ & $\gamma$\\
   (1)   &  (2)   & (3) &  (4)   &      (5)      &  (6)&  (7)  & (8) & (9) & (10)\\
  \hline
  All & LS           & 232  & 0.90 ($<$$10^{-5}$)     & 0.42 ($<$$10^{-5}$) &  &  -            & $0.44\pm0.01$ & $1.01\pm0.01$ \\
  All & Bayesian     & 232  & $0.918_{-0.012}^{+0.010}$ & & 0.38 & $0.350_{-0.018}^{+0.020}$ & $0.302_{-0.025}^{+0.026}$ & $0.986_{-0.030}^{+0.030}$  & $2.003_{-0.113}^{+0.120}$ \\
\\
  Type 1 & LS           & 137  & 0.88 ($<$0.0013)  & 0.40 ($<$$10^{-5}$) &   &  -            & $0.52\pm0.01$ & $0.89\pm0.01$ \\
  Type 1 & Bayesian     & 137  & $0.904_{-0.020}^{+0.017}$ & & 0.40 & $0.353_{-0.022}^{+0.028}$ & $0.377_{-0.034}^{+0.033}$ & $0.941_{-0.041}^{+0.042}$ &  $2.372_{-0.175}^{+0.197}$  \\
\\
  Type 2 & LS           &  95  & 0.88 ($<$$10^{-5}$)  & 0.45 ($<$$10^{-5}$) &   &  -            & $0.24\pm0.01$ & $1.00\pm0.01$  \\
  Type 2 & Bayesian     &  95  & $0.889_{-0.027}^{+0.023}$ & & 0.40 & $0.325_{-0.027}^{+0.031}$ & $0.180_{-0.047}^{+0.048}$ & $0.926_{-0.059}^{+0.058}$ &  $1.513_{-0.158}^{+0.180}$    \\
  \hline
 \end{tabular}

$Notes$. Column 1: sample; Column 2: linear regression technique. LS:
 least squares; Column 3: number of objects; Column 4: linear
 correlation coefficient. For least squares the Spearman's rank and
 the associated probability that the correlation is entirely caused by
 distance based on our Monte Carlo simulations; Column 5: Kendall's
 partial correlation coefficient and probability that the correlation
 is entirely caused by distance based on our Monte Carlo simulations;
 Column 6: observed scatter in $L_{\rm 6 \mu m}$ about the regression
 line; Column 7: intrinsic scatter about the regression line. At a
 given X-ray luminosity, $\sigma_{\rm int}$ denotes the standard
 deviation in the mid-IR luminosity; Columns 8 and 9: intercept and
 slope of the regression, respectively (Equation~\ref{eq1}); Column
 10: constant in linear space (Equation~\ref{eq2}). Uncertainties are
 1$\sigma$.
 \end{table*}

If we assume instead that the hosts of our AGN with $z$$>$0.5 are
Ultraluminous Infrared Galaxies (ULIRGs) with $L_{\rm 8-1000\,\mu
  m}$$\leq$${\rm 6\times 10^{12}\,}L_\odot$ (very IR-luminous galaxies
become more numerous at high $z$), and SFRs$\sim$1000 $M_\odot {\rm
  yr^{-1}}$, we find that the contamination from star formation at
rest-frame 6\,\mic is less than 15 per cent for 89.1 per cent of these
sources. Table~\ref{tab01} lists our (very conservative) median upper
limits on the level of contamination from star formation at rest-frame
6\,\mic\, as a function of AGN class and luminosity.

For comparison, the median upper limit on the star formation
contamination to the \wise\, 12\,\mic\, passband is 18 per cent at AGN
luminosities $10^{42}$$<$$L_{\rm 2-10\,keV}$$<$$10^{43}\lum$, and less
than 4 per cent at higher luminosities. At the rest-frame wavelengths
sampled with the \wise\, 3.4 and 4.6\,\mic\, passbands the
contamination from star fromation should be negligible.

We note that to compute the above numbers we have used the full sample
of objects i.e., including those for which we have only upper limits
on their 22\,\mic\, fluxes. From the comparison of the 12\,\mic\, flux
distributions and ($f_{\rm 22}/f_{\rm 12}$) IR flux ratios (using
1$\sigma$ upper limits for $f_{\rm 22}$ if not detected), we have
found that the objects in BUXS that have escaped detection at
22\,\mic\, are $\sim$4-5 times fainter at 12\,\mic\, than those
detected at 22\,\mic, but they also have on average lower
($\sim$0.8-0.9 times smaller) IR flux ratios. This implies even lower
levels of contamination from dust in star forming regions in the
objects not detected with \swise at 22\,\mic.

\subsection[]{Reliability of rest-frame 6\,\mic\, luminosities for $z$$>$1 AGN}
\label{l6_highz}
It is well known that there is some curvature in the IR continuum
emission of AGN at rest-frame wavelengths close to the $\sim$1\,\mic\,
inflection point between the UV and near-IR bumps (\citealt{elvis94};
\citealt{richards06}). This means that our linear
interpolation/extrapolation approach, based on the \swise fluxes at
4.6 and 12\,\mic, might be overestimating the 6\,\mic\, luminosities
for objects at $z$$>$1, when the \swise 4.6\,\mic\, band is sampling
rest-frame wavelengths $\lesssim$2\,\mic. To investigate such effect
we have compared our derived 6\,\mic\, luminosities with those
obtained using the \swise fluxes at 12 and 22\,\mic. We have used for
this comparison all AGN in BUXS with $z$$>$1 and detection with
SNR$\geq$2 at 22\,\mic\, (19 out of 37 objects with $z$$>$1). We have
found a median (mean) difference in the logarithm of the 6\,\mic\,
luminosities of just -0.02 (-0.02). Although the numbers involved in
the comparison are small, these results show that at up to $z$$\sim$2,
rest-frame 6\,\mic\, AGN luminosities can be robustly determined using
the \swise 4.6 and 12\,\mic\, fluxes alone. In addition, we confirm
that contamination from dust emission in star forming regions to the
\swise 22\,\mic\, fluxes is not important.

\begin{figure*}
  \centering
  \begin{tabular}{cc}
    \hspace{-0.7cm}\includegraphics[angle=90,width=0.75\textwidth]{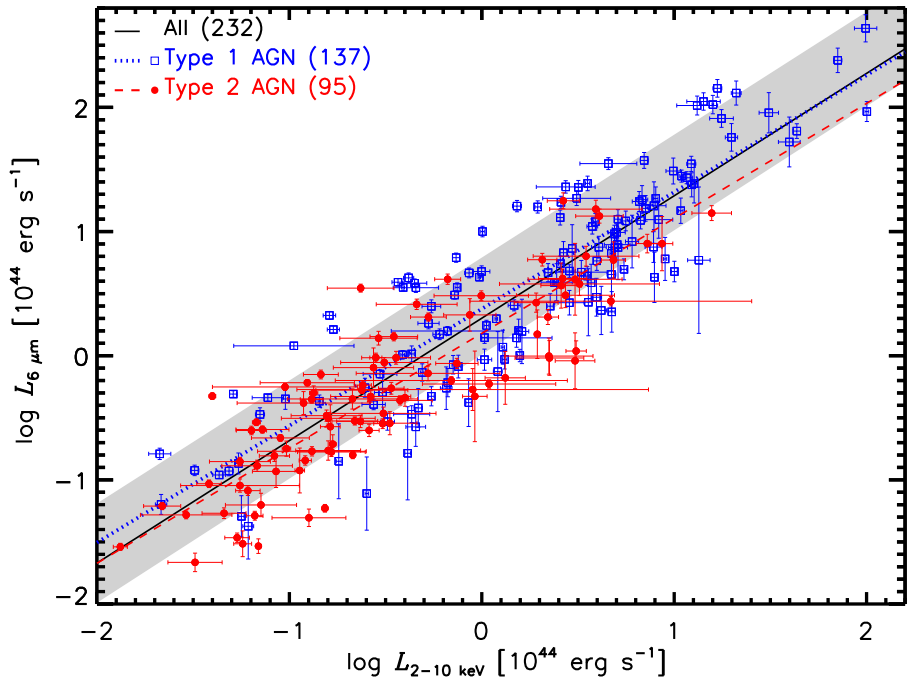}
  \end{tabular}
  \caption{Rest-frame 6\,\mic\, monochromatic luminosities against
    2-10 keV intrinsic (i.e., corrected for absorption) X-ray
    luminosities. The 6\,\mic\, luminosities have been corrected for
    the accretion disk emission and contamination from the AGN host
    galaxies as indicated in Sec.~\ref{l6}. The best-fit linear
    relations in log-log space derived with the K07 Bayesian approach
    for the full sample and the type 1 and type 2 AGN are indicated
    with solid, dotted and dashed lines, respectively. We compare our
    results with the relation derived for type 1 AGN by \citet{lutz04}
    (grey shading).}
  \label{fig5}
\end{figure*}

\section[]{Correlation of $L_{\rm 6\,\lowercase{\mu m}}$ and $L_{\rm {2-10\,\lowercase {ke}V}}$ luminosities}
\subsection[]{Linear regression technique}
\label{technique}
From now own, we will refer to the mid-IR monochromatic luminosities
at rest-frame 6\,\mic\, as $L_{\rm 6\,\mu m}$ while $L_{\rm
  2-10\,\lowercase {ke}V}$ are the rest-frame 2-10 keV X-ray
luminosities corrected for X-ray absorption (see Sec.~\ref{xspec}).
To determine the relationship between $L_{\rm 6\,\mu m}$ and $L_{\rm
  2-10\,keV}$ we have performed a linear regression in log-log space
using two different approaches: the least squares $\chi^2$
minimization \citep{press92} and the Bayesian maximum likelihood
method proposed by \citet[hereafter
  K07]{kelly07}\footnote{http://idlastro.gsfc.nasa.gov/ftp/pro/math/linmix\_err.pro}. For
the K07 technique the data are fitted using the Metropolis-Hastings
Markov Chain Monte Carlo (MCMC) algorithm sampler. We have adopted
uniform prior distributions for the regression parameters. The
best-fit parameters and their corresponding uncertainties have been
determined by taking the median and the 16 and 84 percentiles (68 per
cent enclosed, equivalent to 1$\sigma$) from the posterior probability
distributions of the model parameters using $10^4$ iterations from the
MCMC sampler. The two regression techniques account for measurement
errors in both the independent and dependent variables. However, since
the K07 approach determines simultaneously both the regression
parameters and the intrinsic scatter\footnote{If $x$ and $y$ are the
  independent and dependent variables, respectively, the scatter in
  $y$ at fixed $x$ that would have been measured if there were no
  errors in $y$.}, it should provide more reliable estimates of the
parameters, their associated uncertainties and of the strength of the
linear correlation. Therefore, from now on we will refer to the
results obtained with the K07 technique, but present those obtained
from the least squares technique for comparison with results in the
literature.

The linear regression techniques used here assume that the measurement
errors are Gaussian. However, this is not the case for either the
X-ray luminosity measurements or the uncertainties associated with the
correction of the mid-IR luminosities for the accretion disk and
starlight emission. To overcome this problem, for each data point with
luminosity $L$ and associated uncertainties ${L^-}$ and ${L^+}$ (all
values in logarithmic units) we have computed a Gaussian function with
mean $L$ and dispersion $\sigma$ such that its integral from $L-L^-$
to $L+L^+$ is equal to the enclosed probability given by the fits
(68.3 per cent for the X-ray luminosity uncertainties and 95.4 per
cent for the mid-IR luminosity uncertainties derived from SEABASs). We
have used the 1$\sigma$ uncertainties derived in this way in our
regression analysis.

To determine the $L_{\rm 6\,\lowercase{\mu m}}$-$L_{\rm
  2-10\,\lowercase {ke}V}$ relationship we have conducted the linear
regression analysis treating $L_{\rm 2-10\,\lowercase {ke}V}$ as the
independent variable and $L_{\rm 6\,\lowercase{\mu m}}$ as the
dependent variable. Our decision is based on the fact that hard
X-rays, produced at distances much closer to the central SMBH
(sub-parsec scales), are a more direct tracer of the AGN intrinsic
power than the mid-IR emission, originating from the dusty torus at
tens of parsec scales\footnote{From a statistical point of view we
  note that, in the presence of intrinsic scatter, the best-fit slopes
  of forward and inverse regressions (i.e., switching the dependent
  and independent variables) can be very different. To circumvent such
  problem both variables are often treated symmetrically using the
  bisector line (e.g. \citealt{isobe90}). However, if the dependent
  variable of the linear regression analysis depends on the parameter
  used to select the sample (although our AGN sample is X-ray flux
  limited we stress that we are only using objects from BUXS with
  $L_{\rm 2-10\,keV}$$>$$10^{42}\,\lum$), the truncation of the
  data set will artificially flatten the measured slope due to the
  increased loss of objects with low X-ray luminosities at low mid-IR
  luminosities (\citealt{kelly07}). Thus, the conventional forward fit
  relation i.e., fitting $L_{\rm 6\,\lowercase{\mu m}}$ at a given
  $L_{\rm 2-10\,\lowercase {ke}V}$, should provide the more accurate
  estimate of the existing relationship between the mid-IR and X-ray
  luminosities for our AGN.}. We have assumed a linear relation of the
form:

\begin{eqnarray}
 {\rm log \left({{\it L}_{6\,\mu m} \over 10^{44}} \right)=\alpha+\beta \times log\left({{\it L}_{2-10\,keV}\over 10^{44}}\right)}
\label{eq1}
\end{eqnarray}
where the luminosities are in units of ${\rm erg\,s^{-1}}$ and
$\alpha$ and $\beta$ are the correlation parameters. The luminosities
have been normalized at $L_{\rm 2-10\,keV}$=$10^{44}\lum$, roughly the
mean AGN power of our sample.

Equation~\ref{eq1} can be written as 
\begin{eqnarray}
 {\rm \left({{\it L}_{6\,\mu m} \over 10^{44}} \right)=\gamma \times \left({{\it L}_{2-10\,keV}\over 10^{44}}\right)^\beta} 
\label{eq2}
\end{eqnarray}
where we have computed $\gamma$ ($\sim$10$^\alpha$) and its 1$\sigma$
uncertainties as the median and the 16 and 84 percentiles of the
probability density distributions of $\gamma$. The latter were
computed with a Monte Carlo sampling of the posterior probability
distributions of $\alpha$.

\begin{figure}
  \centering
  \begin{tabular}{cc}
    \hspace{-0.7cm}\includegraphics[angle=90,width=0.496\textwidth]{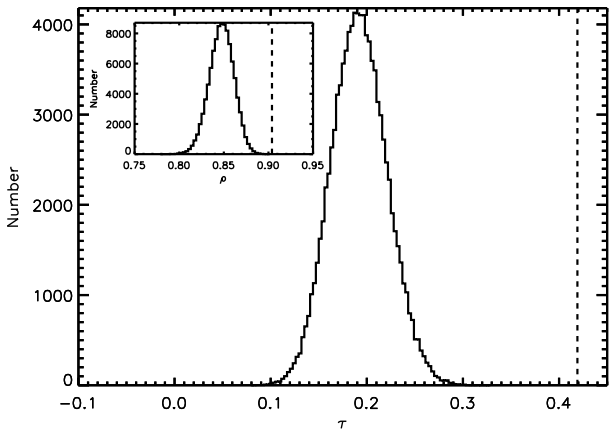}
  \end{tabular}
  \caption{Results of the Monte Carlo simulations conducted to
    determine the correlation between $L_{\rm 6\,\mu m}$ and $L_{\rm
      2-10\,\lowercase {ke}V}$ (solid histograms). The large histogram
    represents the distribution of the Kendall partial correlation
    coefficient (where distance effects have been taken into account)
    while the inset plot illustrates the results for the Spearman
    correlation coefficient. The vertical dashed lines represent the
    correlation coefficients obtained for the original sample.}
  \label{fig5a}
\end{figure}

\subsection[]{The $L_{\rm 6\,\lowercase{\mu m}}$-$L_{\rm 2-10\,\lowercase {ke}V}$ relation}
\label{lcorr}
Table~\ref{tab1} summarizes the results of our regression
analysis. The best-fit linear relations for the full sample and the
type 1 and type 2 AGN obtained with the K07 technique are illustrated
in Fig.~\ref{fig5}. Table~\ref{tab_Appendix} lists the X-ray and
6\,\mic\, luminosities of our AGN.

Although the K07 technique determines the strength of the correlation
between $L_{\rm 6\,\mu m}$ and $L_{\rm 2-10\,\lowercase {ke}V}$, we
have also computed the Spearman rank-order correlation coefficient
($\rho$) for comparison. Both the Spearman's rank and the Bayesian
correlation coefficients confirm the tight correlation between mid-IR
and X-ray luminosities over more than three decades in AGN luminosity
(from $10^{42}-10^{46}\,\lum$).

To investigate the role of distance in our luminosity-luminosity
correlation, we have also performed a partial Kendall $\tau$
correlation test taking into account the common dependence of $L_{\rm
  6\,\mu m}$ and $L_{\rm 2-10\,\lowercase {ke}V}$ on the distance. To
compute the significance of the measured Spearman $\rho$ and Kendall
$\tau$ correlation coefficients for our sample we have conducted a
"scrambling test" (e.g. \citealt{merloni06}; \citealt{bianchi09}). The
idea behind this test is as follows: for each source we keep its X-ray
luminosity and $z$. We then assign them 6\,\mic\, fluxes from the
  sample at random with replacement (bootstrap) and they are converted
  to a mid-IR luminosity using the $z$ of the source. In this way we
are removing any intrinsic correlation between $L_{\rm 6\,\mu m}$ and
$L_{\rm 2-10\,\lowercase {ke}V}$ that is not associated to distance
effects. We have repeated this Monte Carlo test 10$^5$ times
calculating each time the Spearman $\rho$ and Kendall $\tau$
correlation coefficients. The simulations were conducted for both the
full sample and for the type 1 and type 2 AGN samples. The results are
listed in Table~\ref{tab1} and are illustrated in Fig.~\ref{fig5a} for
the full sample (similar results were obtained for the type 1 and type
2 AGN). The values of the correlation coefficients obtained for the
original dataset are well above the distribution of those for the
simulated samples. Our analysis shows that the probability that the
measured correlation between $L_{\rm 6\,\mu m}$ and $L_{\rm
  2-10\,\lowercase {ke}V}$ is entirely due to the range of distances
in our sample is almost always $<$$10^{-5}$ (see Table~\ref{tab1}).

At a given X-ray luminosity, the standard deviation in 6\,\mic\,
luminosities is very similar for both type 1 ($\sim$0.353 dex) and
type 2 ($\sim$0.325 dex) AGN and the slopes of the linear regression
relations are compatible, within the uncertainties. Despite this, type
1 AGN are overall $\sim$1.3-2 times brighter than type 2 AGN at
6\,\mic\, at a given X-ray luminosity, albeit with a large overlap
between both populations.

The shaded area in Fig.~\ref{fig5} shows the relation (and 1$\sigma$
dispersion) for type 1 AGN from \citet{lutz04}. We have compared the
mid-IR/X-ray luminosity ratio distributions of our AGN and the 42
objects in \citet{lutz04} with $L_{\rm 2-10\,keV}$$>$$10^{42}\lum$,
detections at 6\,\mic\, and estimates of the intrinsic X-ray
luminosity. According to the two-sample Kolmogorov-Smirnov (KS) test,
the probability of rejecting the null hypothesis (the two samples are
drawn from the same parent population) is 61.3 per cent. We can
therefore conclude that the two distributions are consistent with each
other. We have also compared the $L_{12\,\mu \rm m}$/$L_{\rm
  2-10\,keV}$ luminosity ratio distributions of our AGN and those from
\citet{gandhi09} with X-ray luminosities $>$$10^{42}\lum$ (37 AGN),
where $L_{12\,\mu \rm m}$ are the monochromatic luminosities in
$\nu$$L_\nu$ units at rest-frame 12\,\mic. To do the comparison we
computed $L_{12\,\mu \rm m}$ (corrected for the accretion disk
emission and contamination from the AGN host galaxies) for each of our
AGN as indicated in Sec.~\ref{l6}. According to the KS test, the
distributions of $L_{12\,\mu \rm m}$/$L_{\rm 2-10\,keV}$ are
consistent with each other (47.3 per cent probability of rejecting the
null hypothesis). Therefore our results are compatible with published
mid-IR:X-ray relations for local AGN (e.g. \citealt{krabbe01};
\citealt{lutz04}; \citealt{gandhi09}; \citealt{levenson09};
\citealt{asmus14}). Nonetheless, they do not support the
luminosity-dependent relationship derived for the type 1 AGN in the
Chandra COSMOS (C-COSMOS) and Chandra Deep Field South (CDFS) surveys
suggested by \citet{fiore09}.

\begin{figure}
  \centering
  \begin{tabular}{cc}
    \hspace{-0.7cm}\includegraphics[angle=90,width=0.496\textwidth]{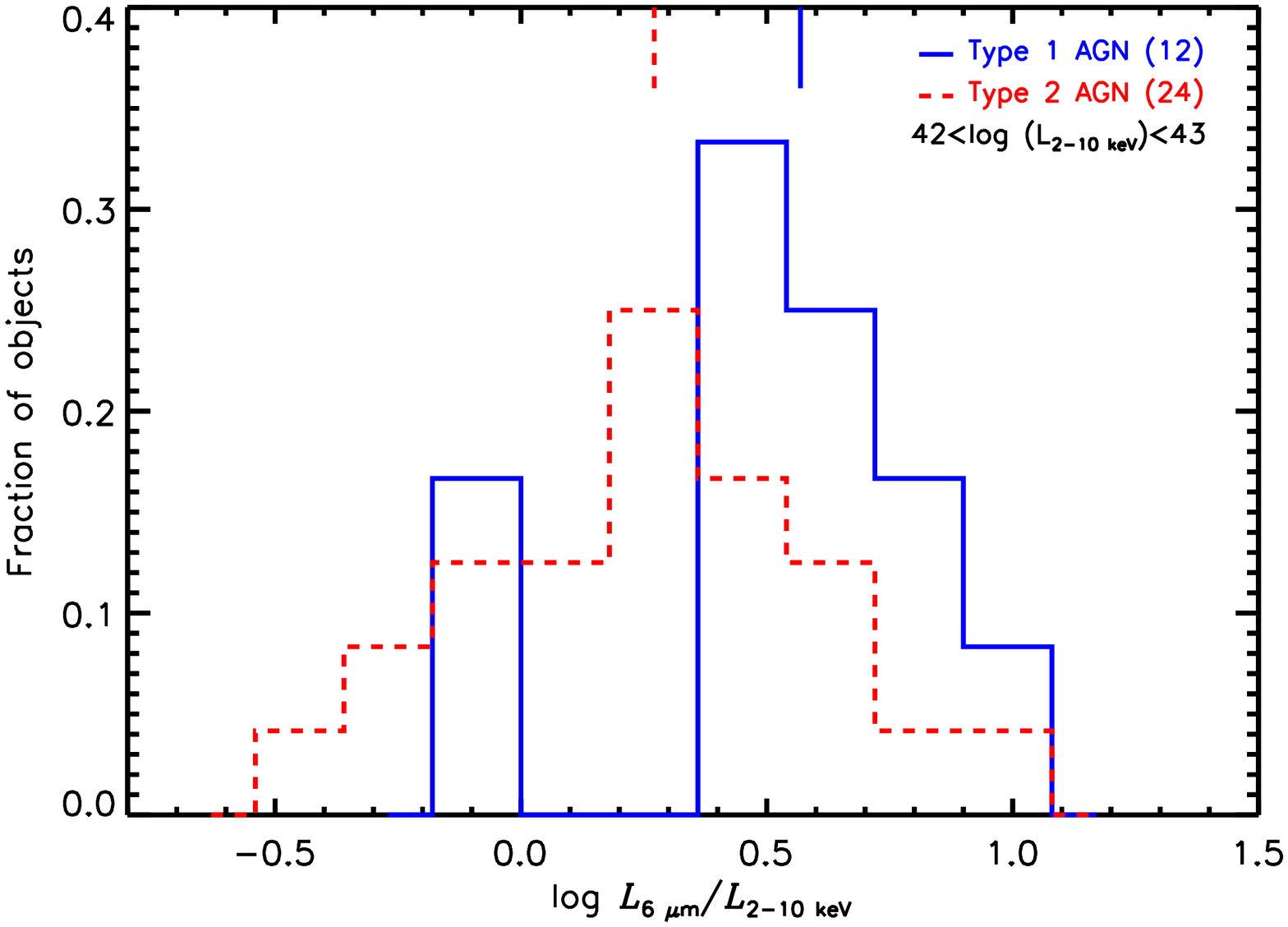}\\
    \hspace{-0.7cm}\includegraphics[angle=90,width=0.496\textwidth]{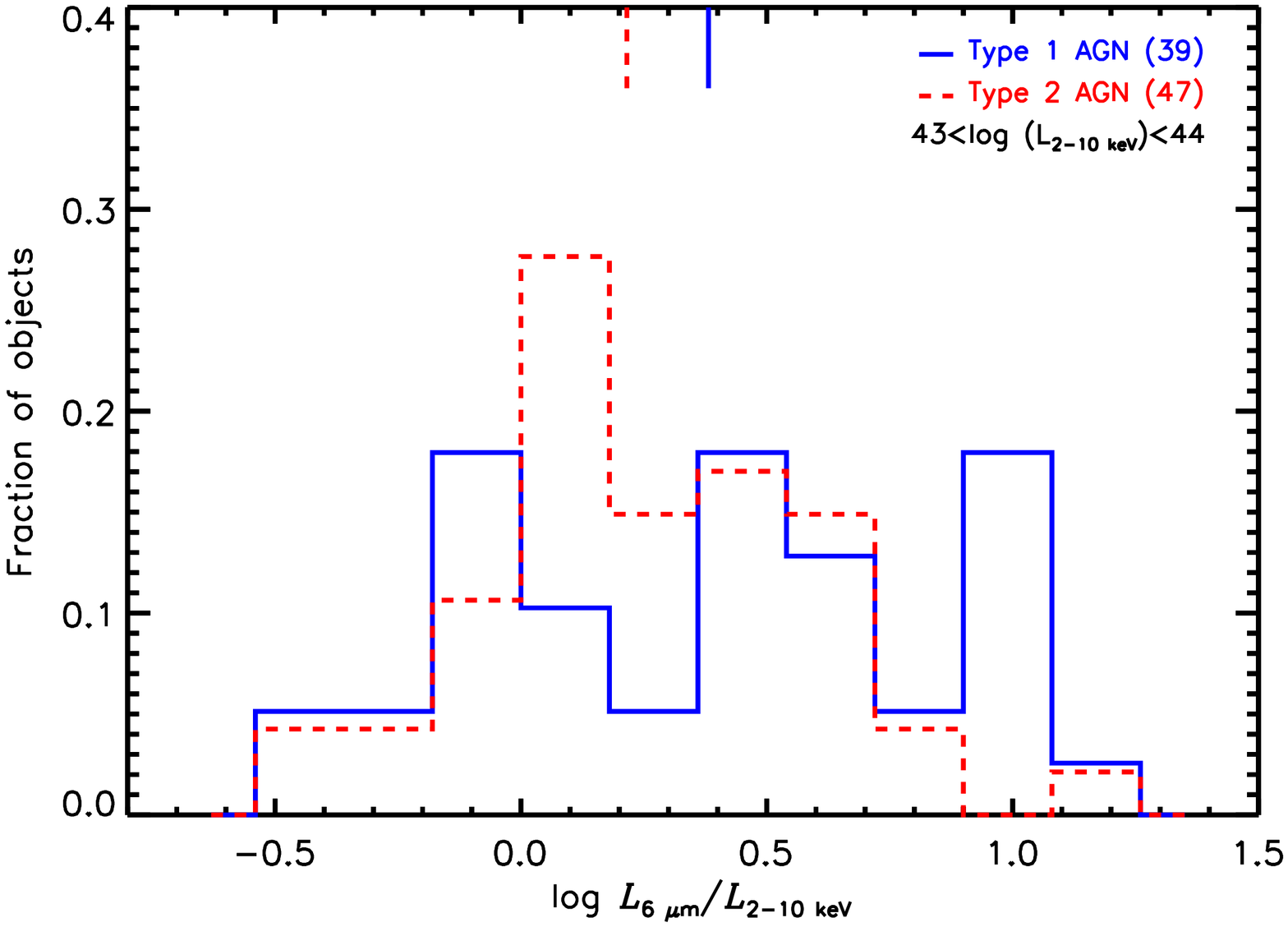}\\
    \hspace{-0.7cm}\includegraphics[angle=90,width=0.496\textwidth]{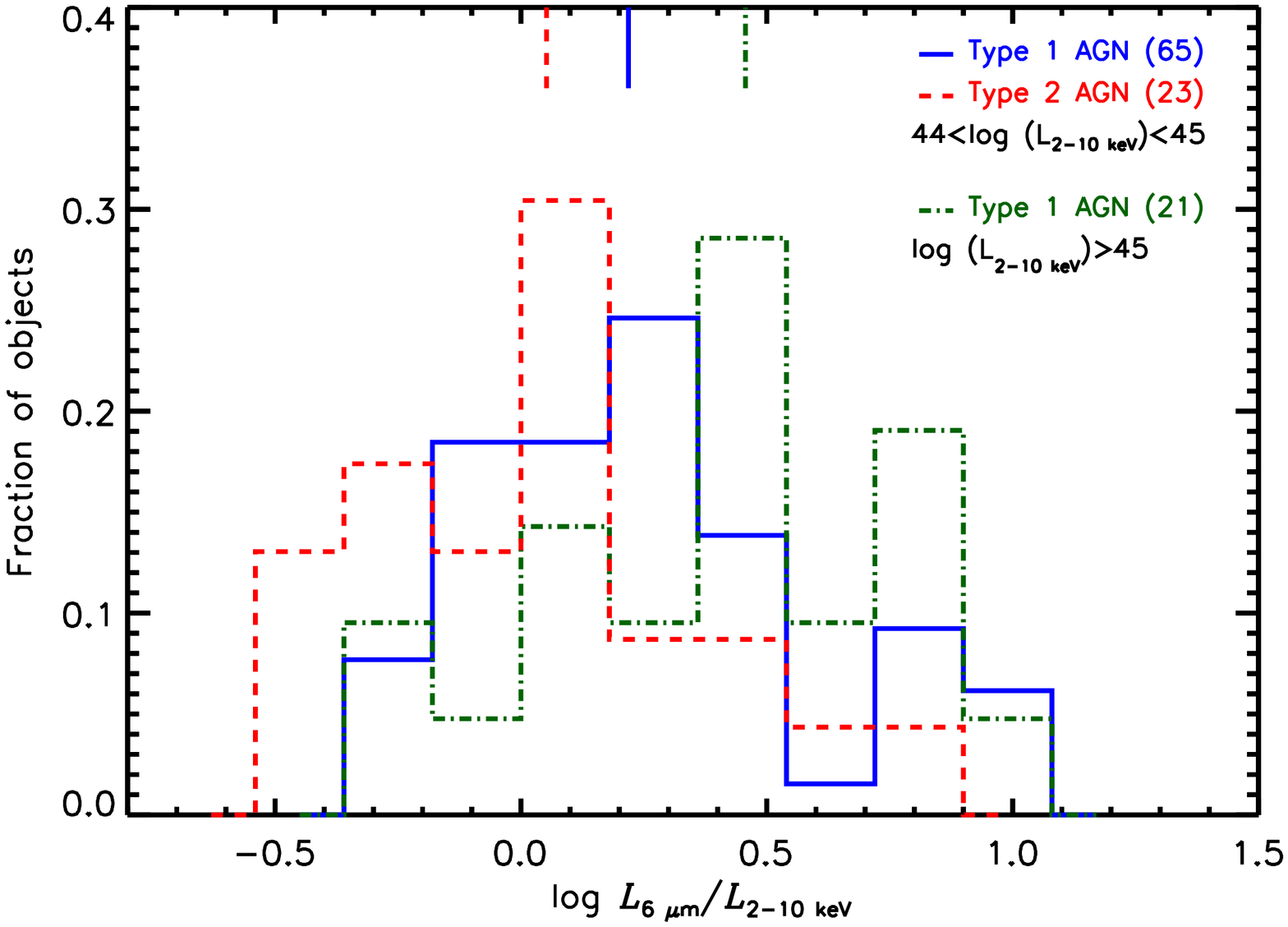}
  \end{tabular}
  \caption{Distribution of 6\,\mic\, to X-ray luminosity ratios (in
    logarithmic units). The vertical lines at the top of the figure
    indicate the median values of the distributions. The bottom
      plot also shows the distribution (and its median) for the type 1
      AGN with $L_{\rm 2-10\,keV}$$>$$10^{45}\lum$}.
  \label{fig6}
\end{figure}

We have demonstrated that at AGN luminosities $L_{\rm
  2-10\,keV}$$>$$10^{42}\lum$, the mid-IR emission detected with
\swise at rest-frame wavelengths $\sim$6\,\mic\, is dominated by hot
dust heated by the AGN. Therefore it is possible to indirectly infer
the intrinsic 2-10 keV X-ray luminosities of AGN with data from \swise
by inverting Equation~\ref{eq1} and Equation~\ref{eq2} after
correcting the 6\,\mic\, luminosities for contamination from the AGN
hosts and the accretion disk. Moreover, we have shown that such
corrections are not negligible, at a $\sim$30 per cent level for AGN
with $10^{42}$$<$$L_{\rm 2-10\,keV}$$<$$10^{43}\lum$ and at a
$\sim$12-18 per cent level for AGN with $10^{43}$$<$$L_{\rm
  2-10\,keV}$$<$$10^{46}\lum$. Based on the full sample, X-ray
luminosities can be determined with a root mean square (rms) error of
0.376 dex. The corresponding rms values for type 1 and type 2 AGN are
0.404 dex and 0.355 dex, respectively\footnote{To determine the rms we
  have used the 6\,\mic\, luminosity range where we can compute the
  dispersion in X-ray luminosities, log\,$L_{\rm 6\,\mu m}$$>$$42.7$
  erg s$^{-1}$ for the full sample and type 2 AGN and log\,$L_{\rm
    6\,\mu m}$$>$$42.8$ erg s$^{-1}$ for type 1 AGN.}.

\begin{table*}
 \caption{Summary of the properties of the distributions of rest-frame 6\,\mic\, over 2-10 keV luminosities.}
 \label{tab4}
 \begin{tabular}{@{}ccccccccccccc}
  \hline
  $\Delta{\rm log({\it L}_{2-10\,keV})}$ & $N^{\rm type\,1}$ & \multicolumn{2}{c} {${\rm log ({\it L}_{6\,\mu m}/{\it L}_{\rm 2-10\,keV})^{\rm type\,1}}$}  & $N^{\rm type\,2}$ & \multicolumn{2}{c} {${\rm log ({\it L}_{6\,\mu m}/{\it L}_{\rm 2-10\,keV})^{\rm type\,2}}$} & & \\
  \cline{3-4} \cline{6-7} 
    ${\rm erg\,s^{-1}}$   &                     & Mean   & Median &   & Mean   & Median & \\
   (1)   &  (2)   & (3) &  (4)   &      (5)      &  (6)&  (7) & & \\
  \hline
  42-45  & 116        & $0.33\pm0.40$ &  $0.31_{-0.38}^{+0.51}$  & 94  & $0.20\pm0.36$  & $0.15_{-0.33}^{+0.42}$ \\ \\ 
  42-43  &  12        & $0.50\pm0.36$ &  $0.57_{-0.61}^{+0.31}$  & 24  & $0.25\pm0.35$  & $0.27_{-0.44}^{+0.32}$  \\
  43-44  &  39        & $0.38\pm0.47$ &  $0.38_{-0.47}^{+0.55}$  & 47  & $0.25\pm0.34$  & $0.21_{-0.25}^{+0.37}$  \\
  44-45  &  65        & $0.26\pm0.34$ &  $0.22_{-0.26}^{+0.49}$  & 23  & $0.04\pm0.36$  & $0.05_{-0.40}^{+0.41}$  \\
  45-46  &  21        & $0.41\pm0.38$ &  $0.46_{-0.33}^{+0.37}$  &  1  &       -         &          -            \\
  \hline
 \end{tabular}

$Notes$. Column 1: X-ray luminosity interval; Column 2: number of type
 1 AGN in the sample; Columns 3 and 4: mean and median of the
 distribution of the mid-IR/X-ray luminosity ratios for type 1 AGN,
 respectively. Column 5: number of type 2 AGN in the sample; Columns 6
 and 7: mean and median of the distribution of the mid-IR/X-ray
 luminosity ratios for type 2 AGN. The quoted
   errors in Columns 3 and 6 are the 1$\sigma$ dispersion of the
   distributions while the errors in Columns 4 and 7 correspond to the
   16 and 84 percentiles.
 \end{table*}

\section{Discussion}
\subsection{Dispersion of the $L_{\rm 6\,\lowercase{\mu m}}$-$L_{\rm 2-10\,\lowercase {ke}V}$ relation} 
Several factors can contribute to the measured scatter of the $L_{\rm
  6\,\mu m}$ versus $L_{\rm 2-10\,keV}$ relationship, such as the
inherent large dispersion in the individual SED shapes associated with
the varying properties of the material responsible for the X-ray and
mid-IR emission and AGN variability effects (the X-ray and mid-IR
observations have been taken several years apart). As X-rays are
produced in a compact region close to the central SMBH, they trace
more directly changes in the AGN bolometric radiative output than the
mid-IR.

AGN torus models with both smooth and clumpy dust distributions
predict anisotropic dust emission at rest-frame
6\,\mic\,(e.g. \citealt{pier92}; \citealt{fritz06};
\citealt{nenkova08a}). Although smooth-density models typically
predict higher levels of mid-IR anisotropy (e.g. $\sim$an order of
magnitude from face-on to edge-on views at rest-frame 6\,\mic;
\citealt{pier92}) it has recently been shown that the anisotropy level
depends mainly on the model assumptions and not so much on the adopted
dust morphology (i.e., smooth or clumpy;
\citealt{feltre12}). Nevertheless, if the mid-IR emission is not
isotropic at the rest-frame wavelengths involved in our study, such
effect should contribute to the measured intrinsic dispersion in the
$L_{\rm 6\,\mu m}$ versus $L_{\rm 2-10\,keV}$ relationship.

We note that, if contamination from star formation were the main
contributor to the observed spread, this would imply a contamination
to the 6\,\mic\, luminosities greater than 1-10$^{-\sigma_{\rm
    int}}$$\sim$56 per cent on average. This is unlikely to be the
case as demonstrated in Sec.~\ref{host_cont_mir}. Moreover, as our
sample consists of AGN with absorption in the Compton-thin regime,
uncertainties associated with the correction of the X-ray luminosities
for intrinsic X-ray absorption are generally small given the good
quality X-ray spectroscopic data available for all sources. This is
supported by the fact that the intrinsic spread of $L_{\rm 6\,\mu m}$
at a given $L_{\rm 2-10\,keV}$ is comparable for (mostly unabsorbed)
type 1 and (highly absorbed) type 2 AGN.

\subsection[]{A luminosity-independent dusty torus?}
\label{ltorus}
Several works in the literature claim that the fraction of AGN
bolometric luminosity that is reprocessed in the mid-IR increases at a
lower rate than the AGN intrinsic bolometric power
(e.g. \citealt{barcons95}; \citealt{maiolino07}; \citealt{treister08};
\citealt{mckernan09}; \citealt{mor09}; \citealt{calderone12}). These
results have been interpreted in terms of a luminosity-dependent
unification model, where the properties of the AGN dusty torus, such
as the geometrical covering factor, vary with luminosity
(e.g. \citealt{lawrence91}; \citealt{simpson05};
\citealt{dellaceca08}). Other studies find, however, either weak or no
luminosity dependence (\citealt{gandhi09}; \citealt{levenson09};
\citealt{roseboom13}).

We have determined a relationship between $L_{\rm 6\,\mu m}$ and
$L_{\rm 2-10\,keV}$ for both type 1 and type 2 AGN that is consistent
with being linear. Furthermore, the average rate of change of $L_{\rm
  6\,\mu m}$ with respect to $L_{\rm 2-10\,keV}$ is similar in the two
AGN classes. Fig.~\ref{fig6} shows the distributions of log($L_{\rm
  6\,\mu m}$/$L_{\rm 2-10\,keV}$) for type 1 and type 2 AGN at
different X-ray luminosities. Table~\ref{tab4} lists the mean and
median of these distributions. We find a mild variation, if at all, of
log($L_{\rm 6\,\mu m}/L_{\rm 2-10\,keV}$) with luminosity, in any case
well within the mutual dispersions.

In this comparison we are implicitly assuming that the 2-10 keV
intrinsic luminosity is a reliable proxy for the AGN bolometric
radiative power and that the hard X-ray bolometric correction is
constant across the luminosity range of our sample. Recent works in
the literature suggest that the X-ray emission of AGN is likely weakly
anisotropic, in the sense that type 2 AGN appear intrinsically fainter
than type 1 AGN at X-ray energies (e.g. \citealt{diamond09};
\citealt{lamassa10}; \citealt{dicken14}; \citealt{liu14} and
references therein). If this was the case, this should have no effect
on the results obtained for our type 1 AGN. On the other hand, taking
into account that the amplitude of the X-ray anisotropy does not seem
to vary with luminosity (\citealt{liu14}) a correction for X-ray
anisotropy effects would move the points for the type 2 AGN in
Fig.~\ref{fig5} to the right but, should not change significantly the
slope of the correlation between their mid-IR and X-ray luminosities.

We note that, if the X-ray bolometric correction increases with AGN
luminosity then the power-law spectral slope of the equivalent
relation between $L_{\rm 6\,\mu m}$ and the bolometric luminosity
$L_{\rm bol}$ would decrease. If we adopt, for example, the
luminosity-dependent bolometric correction of \citet{marconi04}, we
would obtain $L_{\rm 6\,\mu m}\propto L_{\rm bol}^{0.750\pm0.024}$ for
the full sample of objects (for a similar luminosity-dependence of the
hard X-ray bolometric correction see also \citealt{hopkins07}). More
recent studies in the literature suggest, however, a much weaker
dependence of the hard X-ray bolometric correction with AGN luminosity
if any (\citealt{vasudevan07}; \citealt{marchese12};
\citealt{fanali13}).

Finally, we do not expect our results to be affected by sample
selection biases. On the one hand, as indicated in
Sec.~\ref{magnitudes}, only six sources in BUXS do not have detections
in AllWISE with SNR above our detection threshold. Hence, our AGN
sample is not biased against infrared-faint sources (these sources
would occupy the lower right corner in Fig.~\ref{fig5}). On the other
hand, spectroscopic follow-up studies of infrared selected AGN
candidates did not find evidence for the existence of a population of
AGN with infrared-to-X-ray luminosity ratios significantly higher than
those of the AGN revealed in X-ray surveys, at least not for AGN with
X-ray luminosities and $z$ in the range sampled with the BUXS survey
(e.g \citealt{lacy13}).

Thus, our study shows that, if the fraction of AGN bolometric
luminosity that is reprocessed in the mid-IR decreases with the AGN
luminosity, the effect is certainly weak in both type 1 and type 2 AGN
over the three decades in X-ray luminosity covered with our
sample. This finding is at odds with simple receding torus models
(e.g. \citealt{simpson05}).

\subsection{Anisotropy of the AGN emission at 6\,\mic}
\label{anisotropy}
\begin{figure}
  \centering
  \begin{tabular}{cc}
    \hspace{-0.7cm}\includegraphics[angle=90,width=0.496\textwidth]{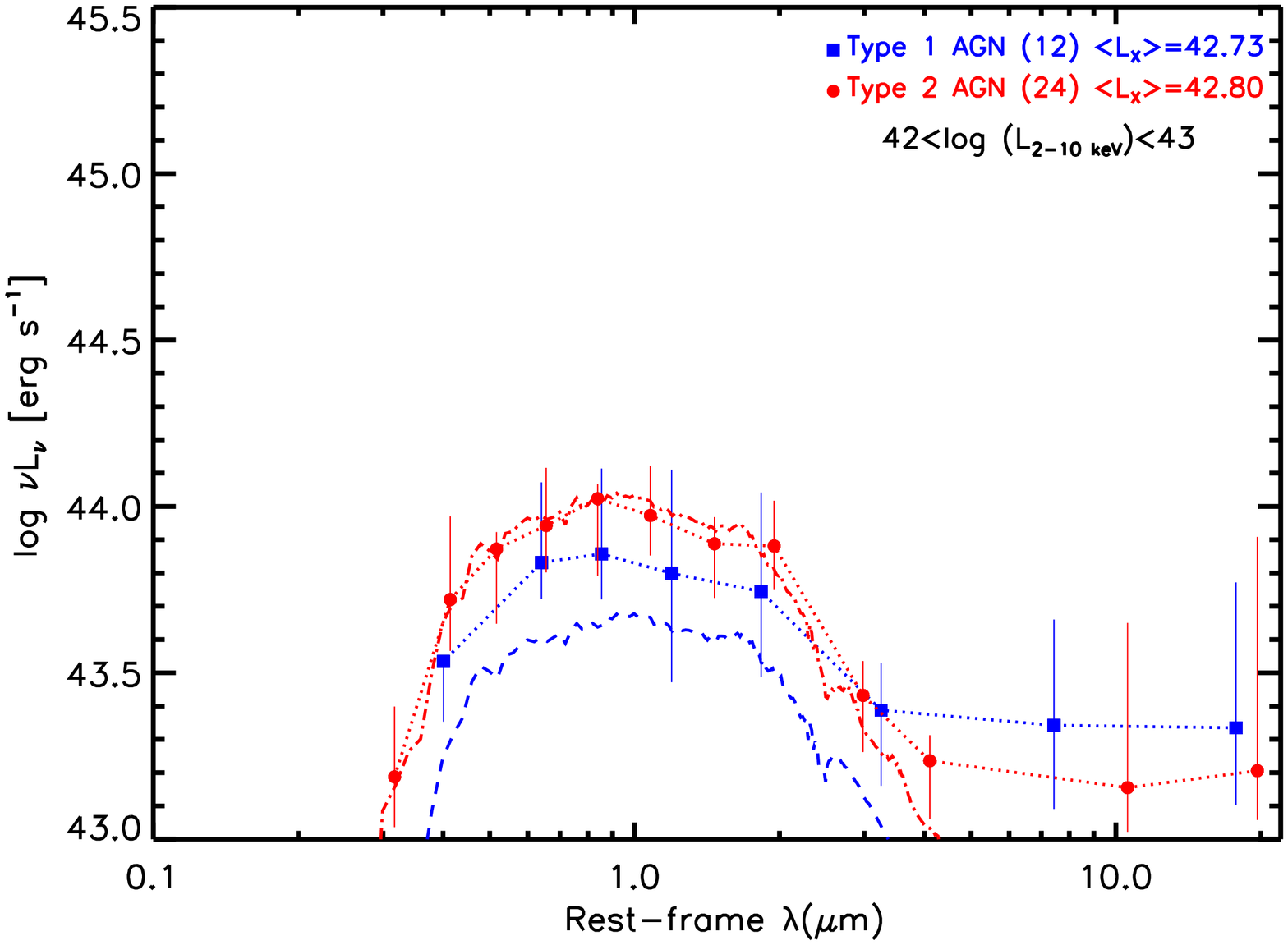}\\
    \hspace{-0.7cm}\includegraphics[angle=90,width=0.496\textwidth]{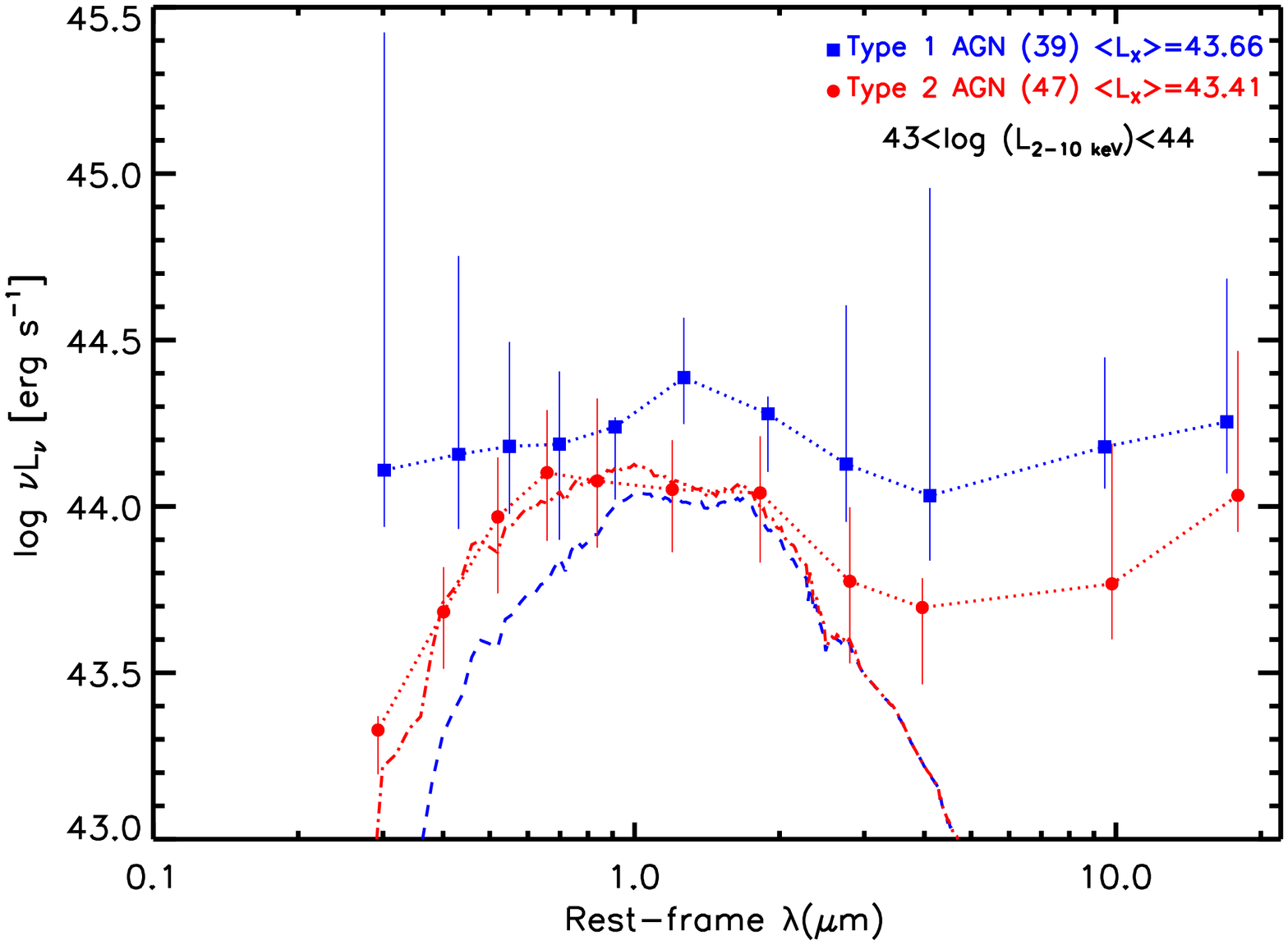}\\
    \hspace{-0.7cm}\includegraphics[angle=90,width=0.496\textwidth]{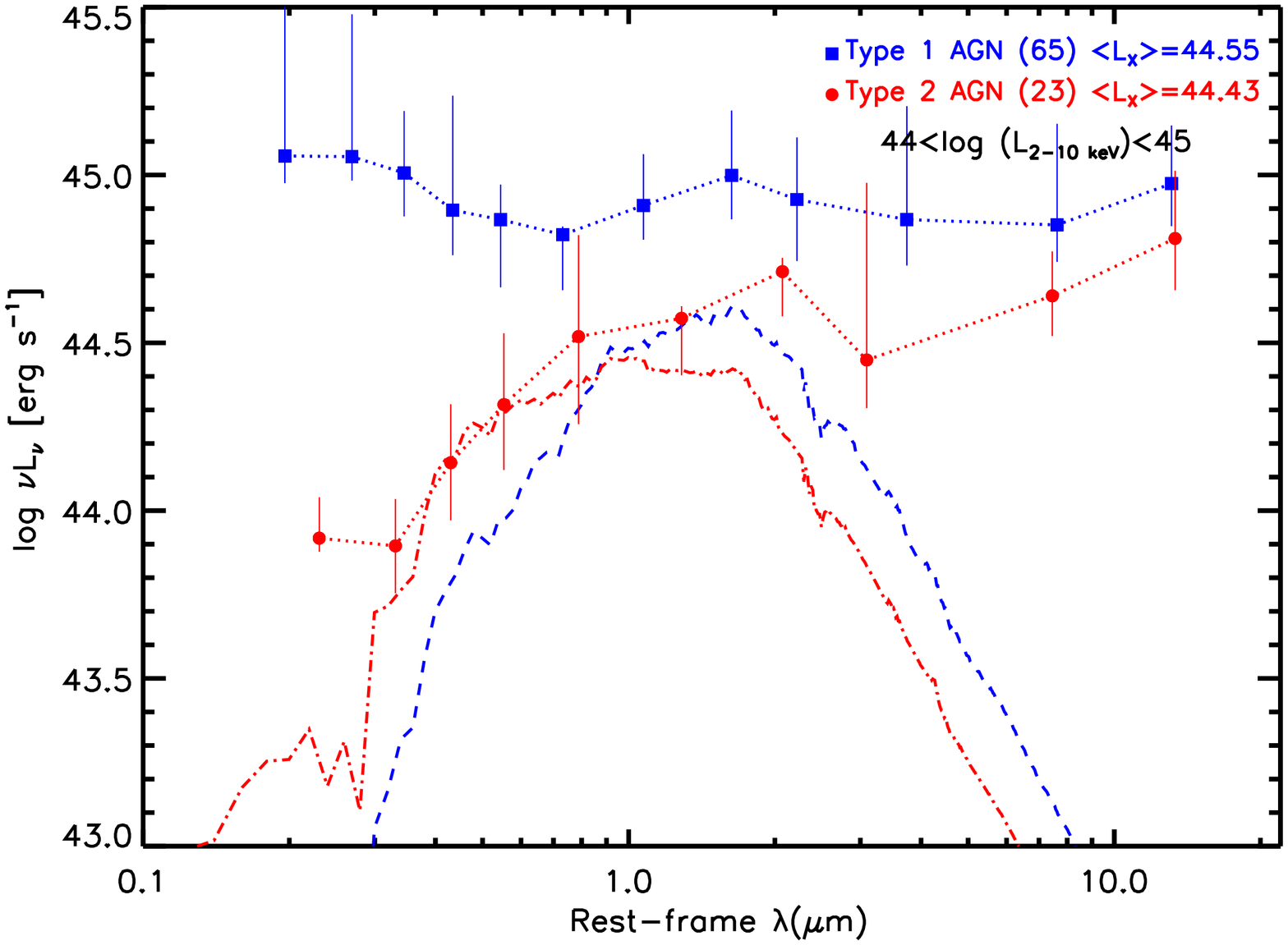}
  \end{tabular}
  \caption{Rest-frame median SEDs of the type 1 and type 2 AGN (filled
    symbols) in three intervals of AGN luminosity before correction
    for host galaxy emission. In the plots are also indicated the mean
    X-ray luminosities of each sub-sample. To help guide the eye the
    points are connected with a power law in linear space (dotted
    lines). The dashed and dotted-dashed lines represent the median of
    the best-fit SEDs of the host galaxies. The vertical error bars
    indicate the 16 and 84 percentiles (68 per cent enclosed,
    equivalent to 1$\sigma$) of the distribution of values.}
  \label{fig7}
\end{figure}

In Sec.~\ref{lcorr} we have shown that at a given X-ray luminosity,
X-ray selected type 1 AGN are on average $\sim$1.3-2 times more
luminous at rest-frame wavelengths $\sim$6\,\mic\, than type 2 AGN
although with a considerable overlap (see e.g. \citealt{ramos07},
\citealt{honig11} and \citealt{yang14} for similar results). A KS test
gives a probability of 2.4 per cent that the distributions of
log($L_{\rm 6\mu m}/L_{\rm 2-10\,keV}$) for type 1 and type 2 AGN are
drawn from the same sample. As pointed out in Sec.~\ref{ltorus}, the
AGN X-ray emission is likely weakly anisotropic. If this is the case,
a correction for such effect would make the 6\,\mic-to-X-ray
luminosity ratios smaller for type 2 AGN and thus the differences in
mid-IR luminosities at a given X-ray luminosity between type 1 and
type 2 AGN would be even larger than our estimation. To better
understand this effect we show in Fig.~\ref{fig7} the median
rest-frame UV-to-mid-IR SEDs of our AGN in three X-ray luminosity bins
before applying any correction for host galaxy contamination. These
were computed by first shifting the individual SEDs to a common
rest-frame wavelength grid. We then converted the fluxes to
luminosities in $\nu$$L$$_\nu$ units and we combined them in
wavelength bins containing at least 15 points. Each value represents
the median of the points in the bin while the dispersion is defined
with the 16 and 84 percentiles. We did not take into account
upper-limits in the computation of the median SEDs hence, as not all
AGN were detected with \swise at 22\,\mic, our median SEDs might be
somewhat redder than they should at the longest wavelengths
($>$10\,\mic).

\begin{figure}
  \centering
  \begin{tabular}{cc}
    \hspace{-0.7cm}\includegraphics[angle=90,width=0.496\textwidth]{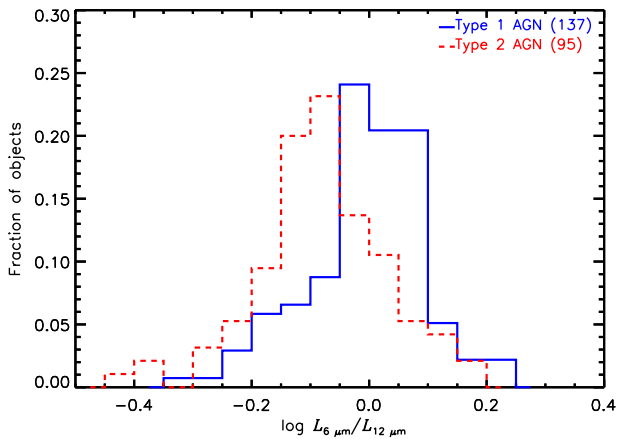}\\
  \end{tabular}
  \caption{Distributions of the rest-frame 6\,\mic\, versus 12\,\mic\,
    mid-IR luminosity ratios for type 1 and type 2 AGN (in logarithmic
    units). The mid-IR luminosities have been corrected for the
    accretion disk emission and contamination from the AGN host
    galaxies.}
  \label{fig8}
\end{figure}

Fig.~\ref{fig7} also shows the median SEDs of the best-fit host galaxy
stellar emission templates. Although a description of the properties
of the host galaxies of our AGN is beyond the scope of this paper, we
see that on average there are no significant differences in the
near-IR luminosities of the type 1 and type 2 AGN hosts (a proxy for
their stellar masses) except at luminosities
$10^{42}-10^{43}\,\lum$. We note, however, that the results obtained
at these luminosities should be treated with caution. The number of
type 1 AGN is rather small and hence the sample might not be
representative of the overall population. Furthermore, at X-ray
luminosities from $10^{42}$ to $10^{43}\lum$, the emission from the
host galaxies could dilute the AGN signatures in a significant
fraction of objects (50 per cent; see Fig. 5 in
\citealt{caccianiga07}). Hence, it is likely that at $L_{\rm
  2-10\,keV}<10^{43}\,\lum$ some type 1 AGN in BUXS might have been
misclassified as type 2 AGN.

The dispersion in the bins illustrated in Fig.~\ref{fig7} reflects not
only changes in the intrinsic continuum shape of the different SED
components between objects but also the varying levels of host galaxy
contamination and the dispersion in AGN luminosities. Fig.~\ref{fig7}
shows that, even after accounting for the small differences in X-ray
luminosities between type 1 and type 2 AGN in the bins, type 2 AGN are
overall fainter sources at rest-frame wavelengths $\gtrsim$4\,\mic\,
out to at least 12\,\mic\, than type 1 AGN at a given X-ray
luminosity.

To investigate in more detail whether the observed continuum shape of
the mid-IR emission associated to the dusty torus in type 1 and type 2
AGN is different we have compared the distributions of the ($L_{6\,\mu
  \rm m}$/$L_{12\,\mu \rm m}$) luminosity ratios for the two AGN
classes. The 12\,\mic\, luminosities were determined using linear
extrapolation in log-log space between the \swise catalogued fluxes at
4.6 and 12\,\mic\, and were corrected for both the accretion disk
emission and contamination from the AGN host galaxies. The results are
presented in Fig.~\ref{fig8}. Clearly, the mid-IR continuum of type 2
AGN is on average redder\footnote{The term red (blue) refers to
  increasing (decreasing) fluxes in $\nu f_\nu$ at longer
  wavelengths.} than for type 1 AGN, albeit with a large overlap
between both populations. According to the KS test, the probability of
rejecting the null hypothesis (the two samples are drawn from the same
parent population) is higher than 99.99 per cent. We note that if we
use instead the observed $L_{6\,\mu \rm m}$ and $L_{12\,\mu \rm m}$
luminosities (i.e., without any correction for contamination) we still
find a probability of rejecting the null hypothesis of 99.6 per
cent. We have not found any significant dependence of log($L_{\rm
  6\,\mu m}/L_{\rm 12\,\mu m}$) on luminosity in either type 1 or type
2 AGN. Furthermore, if we split the sample into luminosity bins as in
Fig.~\ref{fig7}, we still can reject the null hypothesis probability
(i.e. that the distributions of log($L_{\rm 6\,\mu m}/L_{\rm 12\,\mu
  m}$) for type 1 and type 2 AGN are drawn from the same parent
population) with a confidence of 99.7 and 99.98 per cent for objects
with $10^{43}$$<$$L_{\rm 2-10\,keV}$$<$$10^{44}\,\lum$ and
$10^{44}$$<$$L_{\rm 2-10\,keV}$$<$$10^{45}\,\lum$, respectively. Using
the KS test, we cannot reject the null hypothesis for objects with
$L_{\rm 2-10\,keV}$$<$$10^{43}\,\lum$, most likely due to the issues
indicated previously.

Our results are in agreement with previous studies of the mid-IR
SEDs of local Seyfert galaxies (e.g. \citealt{alonso-herrero03};
\citealt{ramos11} and references therein) and luminous quasars at
intermediate $z$ (e.g. \citealt{hernan09}).

\subsection{Dependence of the rest-frame 6\,\mic\, emission on the X-ray absorption?}
\label{dep_nh}
As expected, the UV-optical dust reddening and line-of-sight X-ray
absorption properties of our AGN are highly correlated, in the sense
that X-ray absorption is less common in objects classified in the
UV/optical as type 1 and their X-ray column densities are on average
lower than those measured for type 2 AGN (e.g. \citealt{mainieri02};
\citealt{mateos05a}; \citealt{tozzi06}; \citealt{winter09};
\citealt{mateos10}; \citealt{corral11} \citealt{scott11};
\citealt{merloni14}). These results have often been interpreted in
terms of orientation-based unified models, where the AGN optical
classification and X-ray column density are tracers of the inclination
angle of the dusty torus with respect to our line-of-sight.

Our study supports the idea that there are differences in the
properties of the dusty torus emission between type 1 and type 2
AGN. What is less clear is whether such emission varies strongly with
the X-ray absorption. For example, the study of \citet{silva04}
suggests that it remains nearly unchanged up to X-ray column densities
approaching the Compton-thick regime but it is significantly different
for Compton-thick AGN.

\begin{figure}
  \centering
  \begin{tabular}{cc}
    \hspace{-0.7cm}\includegraphics[angle=90,width=0.496\textwidth]{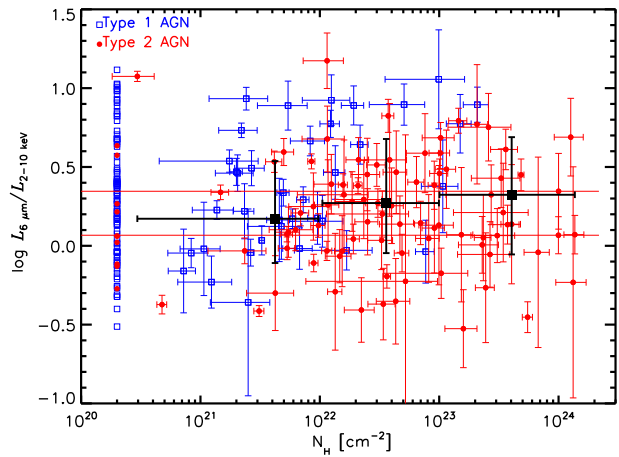}
  \end{tabular}
  \caption{Mid-IR to X-ray luminosity ratio (in logarithmic units)
    versus the line-of-sight X-ray absorption. For clarity, we have
    arbitrarily assigned a column density of ${\rm
      2\times10^{20}\,cm^{-2}}$ to the unabsorbed objects. Large
    filled squares and their vertical error bars are the median and 16
    and 84 percentiles in different column density intervals indicated
    with the horizontal error bars.}
  \label{fig9}
\end{figure}

To investigate whether the dusty torus emission depends on the X-ray
absorption we have computed the strength of the correlation between
log($L_{\rm 6\,\mu m}$/$L_{\rm 2-10\,keV}$) and ${\rm N_H}$ in our
sample of AGN, where $L_{\rm 6\,\mu m}$ has been corrected for the
emission from the accretion disk and the AGN hosts as indicated in
Sec.~\ref{decomp}. Based on the F-test probability at a 95 per cent
confidence level 37 type 1 AGN and 85 type 2 AGN are X-ray absorbed
(see Sec.~\ref{xspec} for details). The data are shown in
Fig.~\ref{fig9}. To determine the strength of the correlation between
log($L_{\rm 6\,\mu m}$/$L_{\rm 2-10\,keV}$) and ${\rm N_H}$ we have
used the K07 technique as it takes into account both the measurement
errors and intrinsic scatter. We have used the full AGN sample
considering the 1$\sigma$ upper limits in ${\rm N_H}$ determined from
the best-fit models for the X-ray unabsorbed objects. 

The distributions of log($L_{\rm 6\,\mu m}$/$L_{\rm 2-10\,keV}$) for
absorbed and unabsorbed objects are indistinguishable (47.0 per cent
probability to reject the null hypothesis). As pointed out before, the
optical classification and X-ray absorption of our AGN are highly
correlated, but the correspondence is not one-to-one. Based on the
F-test probability at a 95 per cent confidence level, 37/137 (27 per
cent) type 1 AGN are X-ray absorbed while 10/95 (10.5 per cent) type 2
AGN are X-ray unabsorbed. If we adopt a column density threshold of
${\rm N_H}$=${\rm 4\times10^{21}\,cm^{-2}}$ to distinguish between
X-ray unobscured and obscured AGN, as in recent studies, the numbers
change to 22/137 type 1 AGN (16 per cent) and 15/95 type 2 AGN (15.8
per cent). This effect smooths out the already small difference in the
$L_{\rm 6\,\mu m}$/$L_{\rm 2-10\,keV}$ luminosity ratio distributions
between type 1 and type 2 AGN.

Moreover, we find that the mid-IR/X-ray luminosity ratio is largely
independent of the line-of-sight X-ray absorption (linear correlation
coefficient $\rho$=0.03$\pm$0.11). Median values of log($L_{\rm 6\,\mu
  m}$/$L_{\rm 2-10\,keV}$) are ${\rm 0.17_{-0.28}^{+0.37}}$ for ${\rm
  N_H<10^{22}\,cm^{-2}}$, ${\rm 0.27_{-0.32}^{+0.41}}$ for ${\rm
  10^{22}<N_H<10^{23}\,cm^{-2}}$ and ${\rm 0.32_{-0.38}^{+0.37}}$ for
${\rm N_H>10^{23}\,cm^{-2}}$ (for similar results see also
\citealt{lutz04}; \citealt{gandhi09}; \citealt{ichikawa12}). This is
an expected result. On the one hand, in X-rays we are measuring the
amount of absorption along our line-of-sight to each AGN. Clearly, the
measured absorption values could change significantly if we were
observing the objects along different lines-of-sight. On the other
hand, at mid-IR wavelengths we are measuring the integrated
reprocessed emission from all the material in the dusty torus in the
plane of the sky. If this material is absorbing the X-rays and it is
re-emitting the radiation almost isotropically, we would expect to
observe approximately the same log($L_{\rm 6\,\mu m}$/$L_{\rm
  2-10\,keV}$) value independently of orientation. We have shown that
the AGN emission at rest-frame 6\,\mic\, is not free of dust
extinction effects. Nevertheless, there is certainly a large overlap
between the log($L_{\rm 6\,\mu m}$/$L_{\rm 12\,\mu m}$) distributions
for the type 1 and type 2 AGN populations ($\sim$0.1 dex of shift; see
Fig.~\ref{fig8}) and the scatter in the luminosity ratio distributions
is significant. This makes orientation-dependent effects, such as a
correlation with the line-of-sight X-ray absorption, difficult to
detect, at least for AGN with X-ray absorption in the Compton-thin
regime (for similar results see also \citealt{silva04};
\citealt{gandhi09}).

To summarize, our results imply that the AGN emission at rest-frame
6\,\mic\, is not completely isotropic, due to self-absorption in the
dusty torus. Such interpretation is consistent with AGN torus models
with both smooth and clumpy dust distributions as they all predict
anisotropic dust emission at rest-frame wavelengths of
6\,\mic\,(e.g. \citealt{pier92}; \citealt{fritz06};
\citealt{nenkova08a}). This effect could be enhanced if, as suggested
by recent results in the literature for local Seyfert galaxies, the
properties of the dusty tori in type 1 and type 2 objects are
different (e.g. \citealt{alonso-herrero11}; \citealt{ramos11}). In
either case, our results imply that AGN surveys at rest-frame
$\sim$6\,\mic\, are subject to modest dust obscuration biases.

\section{Conclusions}
We investigate whether rest-frame 6\,\mic\, mid-IR continuum
luminosities associated to the dusty torus emission of AGN determined
with data from the Wide-field Infrared Survey Explorer (\wise) are a
reliable and isotropic proxy of the AGN intrinsic luminosity. To do so
we have studied the correspondence between the rest-frame 6\,\mic\,
and intrinsic 2-10 keV continuum luminosities of AGN using a complete,
X-ray flux limited sample of 137 type 1 AGN and 95 type 2 AGN drawn
from the Bright Ultra-hard XMM-Newton Survey. The AGN in this study
have 2-10 keV X-ray luminosities between $10^{42}$ and $10^{46}{\rm
  \,erg\,s^{-1}}$, $z$ in the range 0.05-2.8 with a median $z$ of
0.558 and Compton-thin X-ray absorption. The X-ray luminosities were
computed from a detailed analysis of the good quality \xmm spectra
available for all objects and are corrected for absorption. To compute
the mid-IR continuum emission associated with the AGN dusty torus at
rest-frame 6\,\mic\, we decomposed the rest-frame UV-to-mid-IR SEDs of
our AGN into AGN and host galaxy components.

To determine the existing relationship between $L_{\rm
  6\,\lowercase{\mu m}}$ and $L_{\rm {2-10\,\lowercase {ke}V}}$
luminosities, we have conducted a linear regression analysis in
log-log space using the Bayesian maximum likelihood technique proposed
by \citet{kelly07}. The main results of our study can be summarized as
follows:
\begin{enumerate}

\item We find that at AGN luminosities $L_{\rm
  2-10\,keV}$$>$$10^{42}\lum$, the mid-IR emission detected with
  \swise at rest-frame $\sim$6\,\mic\, is dominated by hot dust heated
  by the AGN. The contamination from the AGN host galaxies and the
  accretion disk at these wavelengths is not negligible, at a $\sim$30
  per cent level for AGN with $10^{42}$$<$$L_{\rm
    2-10\,keV}$$<$$10^{43}\lum$ and at a $\sim$12-18 per cent level
  for AGN with $10^{43}$$<$$L_{\rm 2-10\,keV}$$<$$10^{46}\lum$.

\item We also found a relationship for the full AGN sample that is
  consistent with being linear, $L_{\rm 6\,\mu m}$\,$\propto$\,$L_{\rm
    2-10\,keV}^{0.99\pm0.03}$, but has significant intrinsic scatter,
  $\Delta$\,log\,$L_{\rm 6\,\mu m}$\,$\sim$\,0.35\,dex. At a given
  X-ray luminosity, both the intrinsic scatter in 6\,\mic\,
  luminosities and the slopes of the linear regression relations are
  compatible, within the uncertainties, for both type 1 and type 2
  AGN.

\item The main contributors to the intrinsic scatter in the $L_{\rm
  6\,\lowercase{\mu m}}$ versus $L_{\rm {2-10\,\lowercase {ke}V}}$
  correlation are likely the well known large dispersion in the
  individual SED shapes associated with the varying properties of the
  material responsible for the X-ray and mid-IR emission and AGN
  variability. If the AGN dust emission at rest-frame wavelengths
  $<$12\,\mic\, is not isotropic, as predicted by models of the AGN
  dusty tori, such effect should also contribute to the measured
  intrinsic dispersion.

\item Assuming both that the 2-10 keV intrinsic luminosity is a
  reliable proxy for the AGN bolometric radiative power and a constant
  X-ray bolometric correction, the fraction of AGN bolometric
  luminosity that is reprocessed in the mid-IR decreases weakly, if at
  all, with the AGN luminosity over the three decades in X-ray
  luminosity covered with our sample. This finding is at odds with
  simple receding torus models.

\item The log($L_{\rm 6\,\mu m}$/$L_{\rm 2-10\,keV}$) luminosity ratio
  is largely independent of the line-of-sight X-ray absorption. If the
  material that absorbs the X-rays re-emits the radiation almost
  isotropically, we should observe approximately the same log($L_{\rm
    6\,\mu m}$/$L_{\rm 2-10\,keV}$) value regardless of orientation,
  yet the X-ray absorption along the line-of-sight to the AGN could
  differ significantly. We have shown that the AGN emission at
  rest-frame 6\,\mic\, is not free of dust extinction
  effects. Nevertheless, taking into account the large overlap between
  the log($L_{\rm 6\,\mu m}$/$L_{\rm 12\,\mu m}$) distributions for
  the type 1 and type 2 AGN populations ($\sim$0.1 dex of shift) and
  the significant scatter in the luminosity ratio distributions, any
  orientation-dependent effects, such as a correlation with the
  line-of-sight X-ray absorption, will certainly be difficult to
  detect, at least for AGN with X-ray absorption in the Compton-thin
  regime.

\item Type 2 AGN have on average redder mid-IR continua and, at a
  given X-ray luminosity, are $\sim$1.3-2 times fainter at rest-frame
  6\,\mic\, than type 1 AGN, although with a considerable overlap
  between both populations. If the AGN X-ray emission is weakly
    anisotropic, as recent results suggest, a correction for such
    effect would increase the differences in mid-IR luminosities at a
    given X-ray luminosity between type 1 and type 2 AGN.
\end{enumerate}

Regardless of whether all AGN have the same or different nuclear dusty
toroidal structures, our results imply that the AGN emission at
rest-frame 6\,\mic\, is not completely isotropic due to self-absorption
in the dusty torus. Thus our results imply that AGN surveys at
rest-frame $\sim$6\,\mic\, are subject to modest dust obscuration
biases.

\section*{Acknowledgments}
This work is based on observations obtained with XMM–Newton, an ESA
science mission with instruments and contributions directly funded by
ESA Member States and NASA. Based on data from the Wide-field Infrared
Survey Explorer, which is a joint project of the University of
California, Los Angeles, and the Jet Propulsion Laboratory/California
Institute of Technology, funded by the National Aeronautics and Space
Administration. Funding for the SDSS and SDSS-II has been provided by
the Alfred P. Sloan Foundation, the Participating Institutions, the
National Science Foundation, the U.S. Department of Energy, the
National Aeronautics and Space Administration, the Japanese
Monbukagakusho, the Max Planck Society and the Higher Education
Funding Council for England. The SDSS Web Site is
http://www.sdss.org/. Based on observations collected at the European
Organization for Astronomical Research in the Southern hemisphere,
Chile, programme IDs 084.A-0828, 086.A-0612, 087.A-0447 and
088.A-0628. Based on observations made with the William Herschel
Telescope and its service programme (programme IDs: SW2009a17,
SW2009b15, SW2010b20)- operated by the Isaac Newton Group, the
Telescopio Nazionale Galileo (programme IDs: TNG12-09B, TNG15-10A,
TNG11-10B, TNG12-11A) - operated by the Centro Galileo Galilei and the
Gran Telescopio de Canarias (programme IDs: GTC38-09B, GTC32-10A,
GTC18-10B, GTC44-11A, GTC41-12A, GTC25-13A) installed in the Spanish
Observatorio del Roque de los Muchachos of the Instituto de
Astrof\'isica de Canarias, in the island of La Palma. SM, FJC, XB,
A.H.-C. and A.A.-H. acknowledge financial support by the Spanish
Ministry of Economy and Competitiveness through grant
AYA2012-31447. SM, FJC and A.A.-H. acknowledge financial support from
the ARCHES project (7th Framework of the European Union,
No. 313146). AA-H acknowledges support from the Universidad de
Cantabria through the Augusto G. Linares program. AC, RDC and PS
acknowledge financial support from the Italian Ministry of Education,
Universities and Research (PRIN2010-2011, grant n. 2010NHBSBE) and
from ASI (grant n. I/088/06/0). The authors wish to thank the
anonymous referee for constructive comments.

\appendix

\section[]{Table of the sample properties}
{\onecolumn
\begin{longtable}{@{\extracolsep{\fill}}lcccc} 
\caption[Properties of the AGN used in this study.]{Properties of the AGN used in this study.} \label{tab_Appendix} \\
  \hline\hline 
  Name & Type & ${\rm log}\,L_{\rm 2-10\,keV}$ & ${\rm log}\,L_{\rm 6\mu m}$\\ [0.5ex] 
   &        & ${\rm erg\,s^{-1}}$ & ${\rm erg\,s^{-1}}$ \\
    \hline 
\endfirsthead

 \caption{continued.}\\
  \hline\hline 
  Name & Type & ${\rm log}\,L_{\rm 2-10\,keV}$ & ${\rm log}\,L_{\rm 6\mu m}$\\ [0.5ex] 
   &        & ${\rm erg\,s^{-1}}$ & ${\rm erg\,s^{-1}}$ \\
   (1)   &  (2)   & (3) &  (4)   \\
    \hline 
\endhead
2XMMiJ000441.2+00071 & 1.9 & 42.76$\pm$0.05 & 42.49$\pm$0.10 \\
2XMMiJ001130.3+00575 &  1 & 45.22$\pm$0.02 & 46.15$\pm$0.07 \\
2XMMJ002146.3-084711 &  1 & 45.00$\pm$0.04 & 45.49$\pm$0.10 \\
2XMMiJ002244.5+00182 &  1 & 43.67$\pm$0.02 & 43.58$\pm$0.16 \\
2XMMJ003430.8-213351 &  1 & 44.21$\pm$0.04 & 44.20$\pm$0.26 \\
2XMMJ004003.8+005853 &  1 & 44.66$\pm$0.15 & 45.55$\pm$0.05 \\
2XMMJ004341.4+005610 &  1 & 44.55$\pm$0.02 & 44.43$\pm$0.27 \\
2XMMJ004404.6+010153 &  2 & 43.48$\pm$0.05 & 43.45$\pm$0.04 \\
2XMMiJ011829.6+00454 &  1 & 43.74$\pm$0.01 & 43.67$\pm$0.08 \\
2XMMJ012447.7+320727 &  1 & 45.06$\pm$0.03 & 45.43$\pm$0.05 \\
2XMMJ013943.0+061254 & 1.5 & 44.46$\pm$0.08 & 44.43$\pm$0.10 \\
2XMMJ014804.3+055055 &  1 & 42.33$\pm$0.06 & 43.21$\pm$0.05 \\
2XMMJ020011.5-093125 &  1 & 44.17$\pm$0.03 & 44.41$\pm$0.04 \\
2XMMJ020112.5-092016 &  2 & 43.33$\pm$0.34 & 43.65$\pm$0.08 \\
2XMMJ022430.4+185315 &  2 & 43.94$\pm$0.14 & 44.33$\pm$0.13 \\
2XMMJ023234.3-073102 &  1 & 44.29$\pm$0.01 & 45.20$\pm$0.05 \\
2XMMJ024200.8+000021 &  1 & 44.84$\pm$0.02 & 45.57$\pm$0.06 \\
2XMMJ024325.4-000412 & 1.5 & 43.48$\pm$0.04 & 43.70$\pm$0.17 \\
2XMMJ030419.5-010911 &  2 & 42.93$\pm$0.12 & 43.07$\pm$0.13 \\
2XMMJ030742.6-000121 & 1.5 & 43.66$\pm$0.05 & 43.43$\pm$0.16 \\
2XMMJ033729.5+004227 &  1 & 43.61$\pm$0.01 & 43.21$\pm$0.38 \\
2XMMiJ044710.5-06255 &  1 & 45.49$\pm$0.05 & 45.96$\pm$0.16 \\
2XMMJ073534.9+435414 & 1.9 & 43.07$\pm$0.35 & 43.62$\pm$0.09 \\
2XMMiJ075734.4+39260 &  2 & 43.38$\pm$0.24 & 43.72$\pm$0.02 \\
2XMMJ080411.1+650906 &  2 & 44.35$\pm$0.18 & 43.98$\pm$0.14 \\
2XMMJ080608.0+244421 &  1 & 43.87$\pm$0.01 & 44.55$\pm$0.04 \\
2XMMJ081014.5+280337 &  1 & 44.83$\pm$0.03 & 45.09$\pm$0.07 \\
2XMMJ082042.4+205715 &  2 & 42.60$\pm$0.02 & 43.67$\pm$0.03 \\
2XMMJ082053.8+210735 &  2 & 44.48$\pm$0.24 & 44.62$\pm$0.29 \\
2XMMJ083049.6+524910 &  1 & 44.67$\pm$0.02 & 44.65$\pm$0.30 \\
2XMMJ083139.1+524205 &  2 & 42.34$\pm$0.10 & 42.79$\pm$0.03 \\
2XMMiJ084012.7+51124 &  1 & 45.99$\pm$0.06 & 46.64$\pm$0.11 \\
2XMMJ084117.5+003448 &  2 & 42.74$\pm$0.31 & 42.95$\pm$0.05 \\
2XMMJ084927.7+445458 &  1 & 44.49$\pm$0.18 & 45.27$\pm$0.06 \\
2XMMJ085228.6+163008 &  2 & 43.54$\pm$0.12 & 44.15$\pm$0.03 \\
2XMMJ085835.2+275543 &  1 & 43.82$\pm$0.29 & 44.20$\pm$0.04 \\
2XMMJ085841.4+140944 &  1 & 45.64$\pm$0.02 & 45.81$\pm$0.06 \\
2XMMJ090053.8+385616 &  2 & 44.04$\pm$0.35 & 43.77$\pm$0.03 \\
2XMMiJ091557.3+29261 &  1 & 44.00$\pm$0.01 & 45.00$\pm$0.04 \\
2XMMiJ091624.3+29391 &  1 & 44.92$\pm$0.07 & 45.10$\pm$0.15 \\
2XMMJ091636.5+301749 &  2 & 42.83$\pm$0.02 & 43.46$\pm$0.03 \\
2XMMJ091645.4+514146 &  2 & 42.98$\pm$0.02 & 43.25$\pm$0.06 \\
2XMMJ092129.2+370103 &  1 & 43.16$\pm$0.01 & 43.63$\pm$0.05 \\
2XMMJ092201.2+301411 &  2 & 44.35$\pm$0.23 & 44.00$\pm$0.15 \\
2XMMJ092313.0+511742 &  1 & 43.82$\pm$0.03 & 43.78$\pm$0.11 \\
2XMMiJ092619.6+36271 &  1 & 44.41$\pm$0.02 & 45.11$\pm$0.06 \\
2XMMJ093347.9+551846 &  1 & 44.75$\pm$0.03 & 45.09$\pm$0.08 \\
2XMMJ093458.2+611234 & 1.9 & 43.44$\pm$0.21 & 43.90$\pm$0.16 \\
2XMMJ094057.1+032401 & 1.5 & 42.51$\pm$0.02 & 43.08$\pm$0.04 \\
2XMMJ094350.2+035913 &  1 & 44.44$\pm$0.15 & 45.36$\pm$0.05 \\
2XMMJ094404.3+480647 &  1 & 43.62$\pm$0.03 & 44.63$\pm$0.04 \\
2XMMJ094439.8+034940 &  1 & 43.23$\pm$0.03 & 44.21$\pm$0.03 \\
2XMMJ094509.0+040817 &  2 & 42.83$\pm$0.18 & 43.11$\pm$0.05 \\
2XMMiJ095630.7-00343 &  1 & 43.64$\pm$0.09 & 43.53$\pm$0.07 \\
2XMMJ095732.0+024302 &  1 & 44.57$\pm$0.02 & 44.59$\pm$0.15 \\
2XMMJ095815.5+014922 &  1 & 45.10$\pm$0.01 & 45.39$\pm$0.16 \\
2XMMJ095857.3+021314 &  1 & 44.95$\pm$0.01 & 44.78$\pm$0.17 \\
2XMMJ095908.3+024309 &  1 & 45.09$\pm$0.02 & 45.38$\pm$0.10 \\
2XMMJ095918.7+020951 &  1 & 44.90$\pm$0.01 & 45.27$\pm$0.09 \\
2XMMJ100015.3+013147 &  2 & 44.68$\pm$0.06 & 44.77$\pm$0.15 \\
2XMMJ100025.2+015852 & 1.5 & 44.01$\pm$0.01 & 44.15$\pm$0.05 \\
2XMMJ100032.1+553630 &  2 & 43.58$\pm$0.17 & 43.64$\pm$0.04 \\
2XMMJ100035.4+052428 & 1.5 & 42.63$\pm$0.03 & 43.04$\pm$0.03 \\
2XMMJ100057.4+684230 &  1 & 44.00$\pm$0.04 & 44.68$\pm$0.05 \\
2XMMJ100120.7+555351 &  1 & 46.00$\pm$0.01 & 45.97$\pm$0.08 \\
2XMMJ100129.3+013633 &  2 & 42.74$\pm$0.16 & 43.15$\pm$0.03 \\
2XMMJ100205.0+023731 &  1 & 44.19$\pm$0.01 & 44.20$\pm$0.09 \\
2XMMJ100237.8+024701 & 1.5 & 43.02$\pm$0.31 & 44.08$\pm$0.03 \\
2XMMJ101616.7+391143 &  1 & 43.74$\pm$0.05 & 44.40$\pm$0.05 \\
2XMMiJ101733.1-00014 & 1.9 & 43.18$\pm$0.02 & 42.77$\pm$0.03 \\
2XMMiJ101811.5+00101 &  1 & 44.71$\pm$0.05 & 44.87$\pm$0.26 \\
2XMMiJ101830.7+00050 &  2 & 42.73$\pm$0.06 & 42.53$\pm$0.04 \\
2XMMJ101922.6+412050 &  1 & 43.51$\pm$0.04 & 43.47$\pm$0.07 \\
2XMMJ102147.4+130850 &  1 & 44.34$\pm$0.02 & 44.67$\pm$0.07 \\
2XMMJ102147.8+131227 & 1.8 & 43.33$\pm$0.01 & 43.20$\pm$0.03 \\
2XMMJ102551.1+384008 & 1.5 & 42.75$\pm$0.01 & 42.71$\pm$0.17 \\
2XMMJ103739.4+414149 &  1 & 44.90$\pm$0.02 & 44.63$\pm$0.20 \\
2XMMJ104048.4+061819 &  1 & 42.79$\pm$0.03 & 42.63$\pm$0.26 \\
2XMMJ104451.4-012226 &  2 & 44.94$\pm$0.06 & 44.90$\pm$0.22 \\
2XMMJ104522.0-012844 &  1 & 44.57$\pm$0.03 & 45.04$\pm$0.06 \\
2XMMJ104522.1+212614 &  1 & 44.55$\pm$0.01 & 45.39$\pm$0.06 \\
2XMMJ104912.6+330501 & 1.9 & 43.42$\pm$0.21 & 43.67$\pm$0.05 \\
2XMMJ105250.0+335505 &  1 & 45.20$\pm$0.00 & 46.02$\pm$0.07 \\
2XMMJ105932.0+242939 &  1 & 44.71$\pm$0.02 & 45.10$\pm$0.08 \\
2XMMiJ111006.8+61252 &  1 & 43.56$\pm$0.03 & 44.59$\pm$0.03 \\
2XMMJ111121.2+482333 &  1 & 44.70$\pm$0.05 & 44.99$\pm$0.07 \\
2XMMJ111121.6+482047 &  1 & 43.65$\pm$0.06 & 44.58$\pm$0.03 \\
2XMMJ111135.6+482945 &  1 & 44.67$\pm$0.01 & 44.82$\pm$0.05 \\
2XMMJ111559.0+425321 &  1 & 44.83$\pm$0.03 & 45.26$\pm$0.11 \\
2XMMJ111606.9+423645 &  1 & 44.55$\pm$0.01 & 44.64$\pm$0.08 \\
2XMMJ111750.7+075710 & 1.5 & 44.74$\pm$0.01 & 44.70$\pm$0.08 \\
2XMMJ111832.4+130732 &  1 & 44.39$\pm$0.08 & 44.59$\pm$0.15 \\
2XMMJ111909.2+130950 &  2 & 44.54$\pm$0.27 & 44.80$\pm$0.08 \\
2XMMJ112026.6+431519 &  2 & 43.12$\pm$0.28 & 43.23$\pm$0.04 \\
2XMMJ112328.0+052823 &  1 & 42.71$\pm$0.02 & 43.69$\pm$0.03 \\
2XMMJ112338.0+052038 &  1 & 45.60$\pm$0.06 & 45.72$\pm$0.20 \\
2XMMJ113121.7+310254 & 1.9 & 43.41$\pm$0.05 & 43.40$\pm$0.05 \\
2XMMJ113129.2+310944 &  1 & 43.26$\pm$0.01 & 43.15$\pm$0.30 \\
2XMMiJ114654.6+20254 &  1 & 44.78$\pm$0.03 & 44.92$\pm$0.21 \\
2XMMJ115754.9+434753 &  2 & 42.58$\pm$0.06 & 42.97$\pm$0.03 \\
2XMMJ120518.6+443926 & 1.9 & 43.60$\pm$0.02 & 43.66$\pm$0.07 \\
2XMMJ120529.5+442106 &  1 & 43.66$\pm$0.12 & 44.55$\pm$0.04 \\
2XMMJ120952.6+393143 &  1 & 42.89$\pm$0.08 & 43.66$\pm$0.03 \\
2XMMJ121118.8+503653 & 1.9 & 43.38$\pm$0.04 & 43.76$\pm$0.02 \\
2XMMJ121122.4+130936 &  1 & 43.86$\pm$0.02 & 44.49$\pm$0.03 \\
2XMMJ121356.1+140431 &  1 & 43.44$\pm$0.06 & 43.61$\pm$0.04 \\
2XMMJ121422.9+024252 &  1 & 45.85$\pm$0.02 & 46.38$\pm$0.10 \\
2XMMJ121509.4+135450 &  1 & 44.50$\pm$0.02 & 45.36$\pm$0.06 \\
2XMMJ121732.7+465829 &  1 & 45.32$\pm$0.03 & 46.12$\pm$0.10 \\
2XMMJ121808.5+471613 &  1 & 43.85$\pm$0.01 & 43.93$\pm$0.07 \\
2XMMJ121839.4+470627 &  2 & 43.09$\pm$0.25 & 43.78$\pm$0.03 \\
2XMMJ121920.6+470323 &  2 & 43.46$\pm$0.20 & 44.14$\pm$0.06 \\
2XMMJ121930.9+064334 &  1 & 42.98$\pm$0.02 & 43.65$\pm$0.08 \\
2XMMJ121952.2+472058 &  1 & 44.41$\pm$0.01 & 44.74$\pm$0.06 \\
2XMMJ122132.3+043557 &  1 & 45.13$\pm$0.06 & 44.77$\pm$0.59 \\
2XMMJ122137.8+043025 & 1.5 & 42.85$\pm$0.02 & 43.53$\pm$0.05 \\
2XMMJ122330.7+154507 &  1 & 42.69$\pm$0.02 & 43.07$\pm$0.04 \\
2XMMJ122532.4+332532 &  1 & 44.11$\pm$0.01 & 44.07$\pm$0.14 \\
2XMMJ122649.5+311735 & 1.9 & 42.82$\pm$0.04 & 42.71$\pm$0.04 \\
2XMMJ122656.4+013124 & 1.8 & 44.61$\pm$0.12 & 45.12$\pm$0.06 \\
2XMMJ123149.0+214749 &  2 & 43.23$\pm$0.02 & 43.29$\pm$0.08 \\
2XMMJ123204.9+215254 &  2 & 45.19$\pm$0.10 & 45.15$\pm$0.06 \\
2XMMJ123305.8+001438 &  1 & 43.40$\pm$0.01 & 42.89$\pm$0.29 \\
2XMMJ123356.1+074755 & 1.5 & 44.12$\pm$0.01 & 43.97$\pm$0.08 \\
2XMMJ123412.7+372734 &  2 & 43.72$\pm$0.09 & 44.31$\pm$0.03 \\
2XMMiJ123602.1+26182 &  2 & 43.53$\pm$0.07 & 43.74$\pm$0.05 \\
2XMMiJ123604.0+26413 &  1 & 43.59$\pm$0.01 & 44.55$\pm$0.03 \\
2XMMJ123625.4+125844 &  2 & 42.78$\pm$0.06 & 42.91$\pm$0.04 \\
2XMMJ123725.2+114158 &  1 & 44.08$\pm$0.02 & 43.87$\pm$0.32 \\
2XMMJ123759.5+621102 &  1 & 44.59$\pm$0.01 & 45.08$\pm$0.07 \\
2XMMJ124127.1+331201 &  2 & 43.96$\pm$0.06 & 43.67$\pm$0.36 \\
2XMMJ124135.8+332624 &  2 & 42.98$\pm$0.38 & 43.75$\pm$0.03 \\
2XMMJ124205.1+323634 &  2 & 44.43$\pm$0.25 & 44.49$\pm$0.10 \\
2XMMJ124213.8-112510 &  1 & 44.42$\pm$0.02 & 44.83$\pm$0.09 \\
2XMMJ124301.0+131217 &  2 & 43.05$\pm$0.02 & 43.08$\pm$0.18 \\
2XMMJ124408.9+113334 & 1.5 & 43.69$\pm$0.01 & 43.86$\pm$0.07 \\
2XMMJ124540.9-002744 &  1 & 45.30$\pm$0.03 & 45.76$\pm$0.11 \\
2XMMJ125304.6+101239 &  2 & 44.67$\pm$0.73 & 44.44$\pm$0.04 \\
2XMMJ125357.1+154314 &  2 & 44.41$\pm$0.11 & 44.56$\pm$0.18 \\
2XMMJ125414.5+101605 &  2 & 42.80$\pm$0.08 & 43.40$\pm$0.03 \\
2XMMJ125453.1+272008 &  2 & 43.37$\pm$0.17 & 44.54$\pm$0.03 \\
2XMMJ125553.0+272405 &  1 & 44.02$\pm$0.01 & 44.24$\pm$0.04 \\
2XMMJ125610.4+260103 &  1 & 44.85$\pm$0.02 & 45.19$\pm$0.10 \\
2XMMJ130237.6-024055 &  2 & 43.12$\pm$0.06 & 43.65$\pm$0.03 \\
2XMMJ130619.0+672421 &  2 & 43.34$\pm$0.12 & 43.47$\pm$0.03 \\
2XMMiJ130906.2+11330 &  2 & 44.59$\pm$0.24 & 45.18$\pm$0.07 \\
2XMMJ130936.2+082815 &  1 & 44.52$\pm$0.13 & 44.67$\pm$0.09 \\
2XMMJ131046.7+271645 &  1 & 44.20$\pm$0.01 & 44.00$\pm$0.05 \\
2XMMiJ131213.6+23195 &  1 & 45.12$\pm$0.11 & 46.02$\pm$0.08 \\
2XMMJ132037.8+341126 &  1 & 42.33$\pm$0.03 & 42.80$\pm$0.08 \\
2XMMJ132101.4+340658 &  1 & 43.78$\pm$0.04 & 44.17$\pm$0.04 \\
2XMMJ132105.4+341500 &  1 & 44.01$\pm$0.04 & 43.97$\pm$0.09 \\
2XMMJ132349.6+654148 &  1 & 43.99$\pm$0.01 & 44.63$\pm$0.03 \\
2XMMJ132419.0+300042 &  2 & 42.12$\pm$0.04 & 42.46$\pm$0.03 \\
2XMMJ132447.4+300900 &  2 & 44.28$\pm$0.11 & 44.43$\pm$0.08 \\
2XMMJ132826.3+583420 &  2 & 43.20$\pm$0.17 & 43.24$\pm$0.08 \\
2XMMJ132958.6+242435 &  1 & 44.36$\pm$0.02 & 44.40$\pm$0.10 \\
2XMMJ133120.3+242304 &  2 & 44.29$\pm$0.02 & 44.17$\pm$0.20 \\
2XMMJ133614.8+520224 &  2 & 44.42$\pm$0.08 & 45.25$\pm$0.06 \\
2XMMJ133811.2+515736 &  2 & 42.92$\pm$0.08 & 43.19$\pm$0.04 \\
2XMMJ134044.5-004516 &  1 & 43.87$\pm$0.02 & 44.79$\pm$0.04 \\
2XMMJ134113.9-005314 &  1 & 44.38$\pm$0.01 & 44.59$\pm$0.03 \\
2XMMJ134133.0-004033 &  2 & 44.31$\pm$0.08 & 44.77$\pm$0.05 \\
2XMMJ134133.1+353252 &  1 & 44.60$\pm$0.02 & 44.76$\pm$0.08 \\
2XMMJ134245.8+403913 &  2 & 42.95$\pm$0.15 & 43.34$\pm$0.03 \\
2XMMJ134252.9+403202 &  1 & 44.69$\pm$0.01 & 44.87$\pm$0.08 \\
2XMMJ134256.5+000057 &  1 & 44.68$\pm$0.02 & 44.35$\pm$0.16 \\
2XMMJ134323.6+001223 &  1 & 44.61$\pm$0.04 & 44.87$\pm$0.09 \\
2XMMJ134511.9+554759 &  1 & 44.82$\pm$0.04 & 45.24$\pm$0.08 \\
2XMMJ134656.6+580316 &  2 & 43.82$\pm$0.07 & 44.62$\pm$0.04 \\
2XMMJ134749.8+582109 &  1 & 45.04$\pm$0.01 & 45.45$\pm$0.05 \\
2XMMJ134803.8-035619 &  2 & 44.51$\pm$0.42 & 44.58$\pm$0.09 \\
2XMMiJ135436.3+05152 &  2 & 42.46$\pm$0.24 & 42.72$\pm$0.03 \\
2XMMiJ135453.5+05071 &  2 & 43.95$\pm$0.91 & 43.73$\pm$0.17 \\
2XMMJ135527.6+181639 &  2 & 44.34$\pm$0.06 & 44.31$\pm$0.06 \\
2XMMJ135628.7+052144 &  2 & 43.49$\pm$0.23 & 43.95$\pm$0.04 \\
2XMMJ140127.6+025606 &  1 & 43.88$\pm$0.01 & 43.91$\pm$0.09 \\
2XMMJ140145.0+025332 &  2 & 43.45$\pm$0.03 & 43.98$\pm$0.04 \\
2XMMJ140248.0+541350 &  2 & 44.49$\pm$0.05 & 44.04$\pm$0.08 \\
2XMMJ140353.8+540939 & 1.9 & 43.52$\pm$0.14 & 43.46$\pm$0.09 \\
2XMMJ140515.4+542459 & 1.9 & 42.51$\pm$0.14 & 42.34$\pm$0.07 \\
2XMMJ140614.2+282340 &  2 & 43.10$\pm$0.19 & 42.69$\pm$0.07 \\
2XMMJ140716.7+281653 &  2 & 43.21$\pm$0.05 & 43.43$\pm$0.18 \\
2XMMJ140719.3+281814 &  1 & 45.09$\pm$0.02 & 45.55$\pm$0.06 \\
2XMMJ140745.3+283028 &  1 & 44.70$\pm$0.07 & 44.89$\pm$0.05 \\
2XMMJ140921.1+261337 &  1 & 44.89$\pm$0.04 & 44.88$\pm$0.12 \\
2XMMJ141449.5+361239 & 1.5 & 42.74$\pm$0.02 & 43.13$\pm$0.08 \\
2XMMJ141512.6+360813 & 1.9 & 43.13$\pm$0.02 & 43.70$\pm$0.06 \\
2XMMJ141531.4+113156 &  1 & 43.59$\pm$0.01 & 44.01$\pm$0.03 \\
2XMMJ141622.4+265631 &  2 & 43.22$\pm$0.18 & 43.22$\pm$0.04 \\
2XMMJ141731.0+265622 &  2 & 43.84$\pm$0.60 & 43.80$\pm$0.03 \\
2XMMiJ142656.4+60290 &  1 & 44.18$\pm$0.02 & 45.20$\pm$0.05 \\
2XMMJ142759.5+262150 &  1 & 44.69$\pm$0.02 & 44.98$\pm$0.07 \\
2XMMJ143025.8+415957 &  1 & 43.72$\pm$0.06 & 44.26$\pm$0.04 \\
2XMMJ143623.8+631726 &  2 & 44.86$\pm$0.06 & 44.90$\pm$0.07 \\
2XMMJ144336.1+062802 &  1 & 44.89$\pm$0.06 & 45.21$\pm$0.19 \\
2XMMJ144404.5+291412 &  1 & 44.41$\pm$0.02 & 45.24$\pm$0.05 \\
2XMMJ144411.3+291508 &  1 & 44.18$\pm$0.18 & 44.14$\pm$0.08 \\
2XMMJ144545.5+292312 &  2 & 43.66$\pm$0.21 & 44.41$\pm$0.04 \\
2XMMJ144936.5+090829 &  1 & 45.03$\pm$0.01 & 45.17$\pm$0.10 \\
2XMMJ145426.6+182956 &  1 & 43.21$\pm$0.03 & 44.32$\pm$0.03 \\
2XMMJ145442.2+182937 &  2 & 42.86$\pm$0.14 & 43.41$\pm$0.03 \\
2XMMJ145459.4+184452 &  1 & 44.46$\pm$0.03 & 44.68$\pm$0.05 \\
2XMMJ145717.5+223332 &  1 & 44.60$\pm$0.03 & 44.47$\pm$0.15 \\
2XMMJ150431.2+474151 &  1 & 45.00$\pm$0.01 & 44.68$\pm$0.08 \\
2XMMJ150558.3+014104 &  2 & 43.20$\pm$0.24 & 43.50$\pm$0.05 \\
2XMMiJ150743.3+01132 &  1 & 45.25$\pm$0.06 & 45.91$\pm$0.07 \\
2XMMJ150930.9+565454 &  1 & 45.15$\pm$0.09 & 46.05$\pm$0.07 \\
2XMMJ151536.7+000347 &  2 & 43.16$\pm$0.09 & 43.85$\pm$0.04 \\
2XMMJ151703.6+562338 & 1.5 & 43.93$\pm$0.01 & 43.62$\pm$0.19 \\
2XMMJ152721.7+360016 &  2 & 43.49$\pm$0.27 & 43.54$\pm$0.05 \\
2XMMJ153202.2+301629 &  2 & 42.84$\pm$0.01 & 42.47$\pm$0.06 \\
2XMMJ153228.8+045358 & 1.5 & 44.08$\pm$0.01 & 44.30$\pm$0.03 \\
2XMMJ153304.0+302508 & 1.8 & 43.19$\pm$0.18 & 43.52$\pm$0.05 \\
2XMMJ154424.1+535546 &  2 & 44.00$\pm$0.25 & 44.48$\pm$0.04 \\
2XMMJ154930.6+213422 &  2 & 44.12$\pm$0.10 & 43.82$\pm$0.21 \\
2XMMiJ155850.7+02353 &  1 & 44.62$\pm$0.03 & 44.36$\pm$0.19 \\
2XMMJ163331.8+570520 &  1 & 43.64$\pm$0.01 & 44.01$\pm$0.06 \\
2XMMJ164119.4+385407 &  1 & 43.39$\pm$0.02 & 43.74$\pm$0.08 \\
2XMMJ164203.4+385300 &  1 & 43.82$\pm$0.03 & 43.74$\pm$0.17 \\
2XMMiJ164723.9+27054 &  2 & 43.87$\pm$0.11 & 43.94$\pm$0.04 \\
2XMMiJ173104.7+36580 & 1.9 & 43.37$\pm$0.12 & 43.47$\pm$0.05 \\
2XMMJ200812.3-111658 &  2 & 44.48$\pm$0.10 & 43.96$\pm$0.22 \\
2XMMJ204043.2-004548 &  2 & 44.42$\pm$0.13 & 44.62$\pm$0.08 \\
2XMMJ204956.3-053457 &  2 & 43.55$\pm$0.18 & 43.98$\pm$0.05 \\
2XMMJ205038.3-053134 &  2 & 42.85$\pm$0.18 & 42.80$\pm$0.09 \\
2XMMJ211516.5+060840 &  1 & 43.93$\pm$0.02 & 44.67$\pm$0.04 \\
2XMMJ212932.8+001044 & 1.9 & 43.08$\pm$0.03 & 43.16$\pm$0.04 \\
2XMMiJ215017.3-05571 &  2 & 42.66$\pm$0.11 & 42.73$\pm$0.04 \\
2XMMiJ215138.7-05224 &  1 & 44.47$\pm$0.02 & 44.86$\pm$0.19 \\
2XMMJ221813.9+001624 &  2 & 43.72$\pm$0.37 & 43.86$\pm$0.06 \\
2XMMiJ232807.6+14420 &  1 & 43.47$\pm$0.09 & 43.85$\pm$0.03 \\
\hline
\end{longtable}
} $Notes$. The targets are listed in order of increasing right
ascention. Column 1: X-ray source name as listed in the second
incremental version of the Second XMM-Newton Serendipitous Source
Catalogue (2XMM-DR3;
http://xmmssc-www.star.le.ac.uk/Catalogue/xcat\_public\_2XMMi-DR3.html);
Column 2: optical spectroscopic classification. $^a$for objects with
a galaxy UV/optical spectrum without emission lines; Column 3: 2-10
keV X-ray luminosity and 1$\sigma$ uncertainty (in logarithmic
units). The luminosities have been corrected for X-ray
absorption. Column 4: rest-frame 6\,\mic\, AGN luminosity associated
to the dusty torus emission and 1$\sigma$ uncertainty (in logarithmic
units). The infrared luminosities have been corrected for both the
accretion disk and host galaxy emission.  \twocolumn \bsp

\label{lastpage}


\begin{thebibliography}{200}

\bibitem[\protect\citeauthoryear{Abazajian et al.}{2009}]{abazajian09}
  Abazajian K.~N., et al., 2009, ApJS, 182, 543

\bibitem[\protect\citeauthoryear{Alonso-Herrero et
    al.}{2011}]{alonso-herrero11} Alonso-Herrero A., et al., 2011,
  ApJ, 736, 82

\bibitem[\protect\citeauthoryear{Alonso-Herrero et
    al.}{2006}]{alonso-herrero06} Alonso-Herrero A., et al., 2006,
  ApJ, 640, 167

\bibitem[\protect\citeauthoryear{Alonso-Herrero et
    al.}{2003}]{alonso-herrero03} Alonso-Herrero A., Quillen A.~C.,
  Rieke G.~H., Ivanov V.~D., Efstathiou A., 2003, AJ, 126, 81

\bibitem[\protect\citeauthoryear{Alonso-Herrero, Ward, \&
    Kotilainen}{1996}]{alonso-herrero96} Alonso-Herrero A., Ward
  M.~J., Kotilainen J.~K., 1996, MNRAS, 278, 902

\bibitem[\protect\citeauthoryear{Antonucci}{1993}]{antonucci93}
  Antonucci R., 1993, ARA\&A, 31, 473

\bibitem[\protect\citeauthoryear{Arnaud}{1996}]{arnaud96} Arnaud
  K.~A., 1996, ASPC, 101, 17

\bibitem[\protect\citeauthoryear{Asmus et al.}{2014}]{asmus14} Asmus
  D., H{\"o}nig S.~F., Gandhi P., Smette A., Duschl W.~J., 2014,
  MNRAS, 439, 1648

\bibitem[\protect\citeauthoryear{Asmus et al.}{2011}]{asmus11} Asmus
  D., Gandhi P., Smette A., H{\"o}nig S.~F., Duschl W.~J., 2011, A\&A,
  536, A36

\bibitem[\protect\citeauthoryear{Assef et al.}{2013}]{assef13} 
Assef R.~J., et al., 2013, ApJ, 772, 26 

\bibitem[\protect\citeauthoryear{Assef et al.}{2010}]{assef10} 
Assef R.~J., et al., 2010, ApJ, 713, 970 

\bibitem[\protect\citeauthoryear{Ballo et al.}{2014}]{ballo14} Ballo
  L., Severgnini P., Della Ceca R., Caccianiga A., Vignali C., Carrera
  F.~J., Corral A., Mateos S., 2014, MNRAS, 444, 2580

\bibitem[\protect\citeauthoryear{Barcons et al.}{1995}]{barcons95}
  Barcons X., Franceschini A., de Zotti G., Danese L., Miyaji T.,
  1995, ApJ, 455, 480

\bibitem[\protect\citeauthoryear{Bianchi et al.}{2009}]{bianchi09}
  Bianchi S., Bonilla N.~F., Guainazzi M., Matt G., Ponti G., 2009,
  A\&A, 501, 915

\bibitem[\protect\citeauthoryear{Bruzual \& Charlot}{2003}]{bruzual03}
  Bruzual G., Charlot S., 2003, MNRAS, 344, 1000

\bibitem[\protect\citeauthoryear{Caccianiga et
    al.}{2008}]{caccianiga08} Caccianiga A., et al., 2008, A\&A, 477,
  735

\bibitem[\protect\citeauthoryear{Caccianiga et
    al.}{2007}]{caccianiga07} Caccianiga A., Severgnini P., Della Ceca
  R., Maccacaro T., Carrera F.~J., Page M.~J., 2007, A\&A, 470, 557

\bibitem[\protect\citeauthoryear{Calderone, Sbarrato, \&
    Ghisellini}{2012}]{calderone12} Calderone G., Sbarrato T.,
  Ghisellini G., 2012, MNRAS, 425, L41

\bibitem[\protect\citeauthoryear{Calzetti et al.}{2000}]{calzetti00}
  Calzetti D., Armus L., Bohlin R.~C., Kinney A.~L., Koornneef J.,
  Storchi-Bergmann T., 2000, ApJ, 533, 682

\bibitem[\protect\citeauthoryear{Cardelli, Clayton, \&
    Mathis}{1989}]{cardelli89} Cardelli J.~A., Clayton G.~C., Mathis
  J.~S., 1989, ApJ, 345, 245

\bibitem[\protect\citeauthoryear{Chabrier}{2003}]{chabrier03} Chabrier
  G., 2003, ApJ, 586, L133

\bibitem[\protect\citeauthoryear{Cohen, Wheaton, \&
    Megeath}{2003}]{cohen03} Cohen M., Wheaton W.~A., Megeath S.~T.,
  2003, AJ, 126, 1090

\bibitem[\protect\citeauthoryear{Corral et al.}{2011}]{corral11}
  Corral A., Della Ceca R., Caccianiga A., Severgnini P., Brunner H.,
  Carrera F.~J., Page M.~J., Schwope A.~D., 2011, A\&A, 530, A42

\bibitem[\protect\citeauthoryear{Cutri et al.}{2013}]{cutri13} Cutri
  R.~M., et al., 2013, Technical Report, Explanatory Supplement to the
  AllWISE Data Release Products

\bibitem[\protect\citeauthoryear{Cutri et al.}{2003}]{cutri03} Cutri
  R.~M., et al., 2003, The IRSA 2MASS All-Sky Point Source Catalog,
  NASA/IPAC Infrared Science
  Archive, available at: http://irsa.ipac.caltech.edu/applications/Gator/

\bibitem[\protect\citeauthoryear{Della Ceca et
    al.}{2008}]{dellaceca08} Della Ceca R., et al., 2008, A\&A, 487,
  119

\bibitem[\protect\citeauthoryear{Diamond-Stanic, Rieke, \&
    Rigby}{2009}]{diamond09} Diamond-Stanic A.~M., Rieke G.~H., Rigby
  J.~R., 2009, ApJ, 698, 623

\bibitem[\protect\citeauthoryear{Dicken et al.}{2014}]{dicken14}
  Dicken D., et al., 2014, ApJ, 788, 98

\bibitem[\protect\citeauthoryear{Dickey \& Lockman}{1990}]{dickey90}
  Dickey J.~M., Lockman F.~J., 1990, ARA\&A, 28, 215

\bibitem[\protect\citeauthoryear{Elbaz et al.}{2011}]{elbaz11} Elbaz
  D., et al., 2011, A\&A, 533, A119

\bibitem[\protect\citeauthoryear{Elvis et al.}{1994}]{elvis94} Elvis
  M., et al., 1994, ApJS, 95, 1

\bibitem[\protect\citeauthoryear{Fanali et al.}{2013}]{fanali13}
  Fanali R., Caccianiga A., Severgnini P., Della Ceca R., Marchese E.,
  Carrera F.~J., Corral A., Mateos S., 2013, MNRAS, 433, 648

\bibitem[\protect\citeauthoryear{Feltre et al.}{2012}]{feltre12}
  Feltre A., Hatziminaoglou E., Fritz J., Franceschini A., 2012,
  MNRAS, 426, 120

\bibitem[\protect\citeauthoryear{Fiore et al.}{2009}]{fiore09} Fiore
  F., et al., 2009, ApJ, 693, 447

\bibitem[\protect\citeauthoryear{Fritz, Franceschini, \&
    Hatziminaoglou}{2006}]{fritz06} Fritz J., Franceschini A.,
  Hatziminaoglou E., 2006, MNRAS, 366, 767

\bibitem[\protect\citeauthoryear{Gandhi et 
al.}{2009}]{gandhi09} Gandhi P., Horst H., Smette A., H{\"o}nig S., Comastri A., Gilli R., Vignali C., Duschl W., 2009, A\&A, 502, 457 

\bibitem[\protect\citeauthoryear{Gordon \& Clayton}{1998}]{gordon98}
  Gordon K.~D., Clayton G.~C., 1998, ApJ, 500, 816

\bibitem[\protect\citeauthoryear{Granato \& Danese}{1994}]{granato94}
  Granato G.~L., Danese L., 1994, MNRAS, 268, 235

\bibitem[\protect\citeauthoryear{Guainazzi, Matt, \&
    Perola}{2005}]{guainazzi05} Guainazzi M., Matt G., Perola G.~C.,
  2005, A\&A, 444, 119

\bibitem[\protect\citeauthoryear{Hern{\'a}n-Caballero et
    al.}{2009}]{hernan09} Hern{\'a}n-Caballero A., et al., 2009,
  MNRAS, 395, 1695

\bibitem[\protect\citeauthoryear{Hewett et al.}{2006}]{hewett06}
  Hewett P.~C., Warren S.~J., Leggett S.~K., Hodgkin S.~T., 2006,
  MNRAS, 367, 454

\bibitem[\protect\citeauthoryear{H{\"o}nig et al.}{2011}]{honig11}
  H{\"o}nig S.~F., Leipski C., Antonucci R., Haas M., 2011, ApJ, 736,
  26

\bibitem[\protect\citeauthoryear{Hopkins, Richards, \&
    Hernquist}{2007}]{hopkins07} Hopkins P.~F., Richards G.~T.,
  Hernquist L., 2007, ApJ, 654, 731

\bibitem[\protect\citeauthoryear{Hopkins et al.}{2004}]{hopkins04}
  Hopkins P.~F., et al., 2004, AJ, 128, 1112

\bibitem[\protect\citeauthoryear{Horst et al.}{2008}]{horst08} Horst
  H., Gandhi P., Smette A., Duschl W.~J., 2008, A\&A, 479, 389

\bibitem[\protect\citeauthoryear{Ichikawa et al.}{2012}]{ichikawa12}
  Ichikawa K., Ueda Y., Terashima Y., Oyabu S., Gandhi P., Matsuta K.,
  Nakagawa T., 2012, ApJ, 754, 45

\bibitem[\protect\citeauthoryear{Isobe et al.}{1990}]{isobe90} Isobe
  T., Feigelson E.~D., Akritas M.~G., Babu G.~J., 1990, ApJ, 364, 104

\bibitem[\protect\citeauthoryear{Jarrett et al.}{2000}]{jarrett00}
  Jarrett T.~H., Chester T., Cutri R., Schneider S., Skrutskie M.,
  Huchra J.~P., 2000, AJ, 119, 2498

\bibitem[\protect\citeauthoryear{Kelly}{2007}]{kelly07} Kelly B.~C.,
  2007, ApJ, 665, 1489

\bibitem[\protect\citeauthoryear{Kennicutt}{1998}]{kennicutt98}
  Kennicutt R.~C., Jr., 1998, ARA\&A, 36, 189

\bibitem[\protect\citeauthoryear{Kotilainen et
    al.}{1992}]{kotilainen92} Kotilainen J.~K., Ward M.~J., Boisson
  C., Depoy D.~L., Smith M.~G., Bryant L.~R., 1992, MNRAS, 256, 125

\bibitem[\protect\citeauthoryear{Krabbe, B{\"o}ker, \&
    Maiolino}{2001}]{krabbe01} Krabbe A., B{\"o}ker T., Maiolino R.,
  2001, ApJ, 557, 626

\bibitem[\protect\citeauthoryear{Lacy et al.}{2013}]{lacy13} Lacy M.,
  et al., 2013, ApJS, 208, 24

\bibitem[\protect\citeauthoryear{LaMassa et al.}{2010}]{lamassa10}
  LaMassa S.~M., Heckman T.~M., Ptak A., Martins L., Wild V.,
  Sonnentrucker P., 2010, ApJ, 720, 786

\bibitem[\protect\citeauthoryear{Lawrence et al.}{2007}]{lawrence07}
  Lawrence A., et al., 2007, MNRAS, 379, 1599

\bibitem[\protect\citeauthoryear{Lawrence}{1991}]{lawrence91} 
Lawrence A., 1991, MNRAS, 252, 586 


\bibitem[\protect\citeauthoryear{Levenson et al.}{2009}]{levenson09}
  Levenson N.~A., Radomski J.~T., Packham C., Mason R.~E., Schaefer
  J.~J., Telesco C.~M., 2009, ApJ, 703, 390

\bibitem[\protect\citeauthoryear{Liu et al.}{2014}]{liu14} Liu T.,
  Wang J.-X., Yang H., Zhu F.-F., Zhou Y.-Y., 2014, ApJ, 783, 106

\bibitem[\protect\citeauthoryear{Lusso et al.}{2013}]{lusso13} Lusso
  E., et al., 2013, ApJ, 777, 86

\bibitem[\protect\citeauthoryear{Lutz et al.}{2004}]{lutz04} Lutz D., Maiolino R., Spoon H.~W.~W., Moorwood A.~F.~M., 2004, A\&A, 418, 465 

\bibitem[\protect\citeauthoryear{Mainieri et al.}{2002}]{mainieri02}
  Mainieri V., Bergeron J., Hasinger G., Lehmann I., Rosati P.,
  Schmidt M., Szokoly G., Della Ceca R., 2002, A\&A, 393, 425

\bibitem[\protect\citeauthoryear{Maiolino et al.}{2007}]{maiolino07}
  Maiolino R., Shemmer O., Imanishi M., Netzer H., Oliva E., Lutz D.,
  Sturm E., 2007, A\&A, 468, 979

\bibitem[\protect\citeauthoryear{Maiolino et al.}{2001}]{maiolino01}
  Maiolino R., Marconi A., Salvati M., Risaliti G., Severgnini P.,
  Oliva E., La Franca F., Vanzi L., 2001, A\&A, 365, 28

\bibitem[\protect\citeauthoryear{Marchese et al.}{2012}]{marchese12}
  Marchese E., Della Ceca R., Caccianiga A., Severgnini P., Corral A.,
  Fanali R., 2012, A\&A, 539, A48

\bibitem[\protect\citeauthoryear{Marconi et al.}{2004}]{marconi04}
  Marconi A., Risaliti G., Gilli R., Hunt L.~K., Maiolino R., Salvati
  M., 2004, MNRAS, 351, 169

\bibitem[\protect\citeauthoryear{Mart{\'{\i}}nez-Sansigre et
    al.}{2005}]{martinez-sansigre05} Mart{\'{\i}}nez-Sansigre A.,
  Rawlings S., Lacy M., Fadda D., Marleau F.~R., Simpson C., Willott
  C.~J., Jarvis M.~J., 2005, Natur, 436, 666

\bibitem[\protect\citeauthoryear{Mason et al.}{2012}]{mason12} Mason
  R.~E., et al., 2012, AJ, 144, 11

\bibitem[\protect\citeauthoryear{Mateos et al.}{2013}]{mateos13} Mateos S., Alonso-Herrero A., Carrera F.~J., Blain A., Severgnini P., Caccianiga A., Ruiz A., 2013, MNRAS, 434, 941 

\bibitem[\protect\citeauthoryear{Mateos et al.}{2012}]{mateos12} Mateos S., et al., 2012, MNRAS, 426, 3271 

\bibitem[\protect\citeauthoryear{Mateos et al.}{2010}]{mateos10}
  Mateos S., et al., 2010, A\&A, 510, A35

\bibitem[\protect\citeauthoryear{Mateos et al.}{2008}]{mateos08}
  Mateos S., et al., 2008, A\&A, 492, 51

\bibitem[\protect\citeauthoryear{Mateos et al.}{2005a,
    2005b}]{mateos05a} Mateos S., Barcons X., Carrera F.~J., Ceballos
  M.~T., Hasinger G., Lehmann I., Fabian A.~C., Streblyanska A.,
  2005b, A\&A, 444, 79

\bibitem[\protect\citeauthoryear{Mateos et
    al.}{2005}]{mateos05} Mateos S., et al., 2005a, A\&A,
  433, 855

\bibitem[\protect\citeauthoryear{Matsuta et al.}{2012}]{matsuta12}
  Matsuta K., et al., 2012, ApJ, 753, 104

\bibitem[\protect\citeauthoryear{McKernan et al.}{2009}]{mckernan09}
  McKernan B., Ford K.~E.~S., Chang N., Reynolds C.~S., 2009, MNRAS,
  394, 491

\bibitem[\protect\citeauthoryear{Merloni et al.}{2014}]{merloni14}
  Merloni A., et al., 2014, MNRAS, 437, 3550

\bibitem[\protect\citeauthoryear{Merloni et al.}{2006}]{merloni06}
  Merloni A., K{\"o}rding E., Heinz S., Markoff S., Di Matteo T.,
  Falcke H., 2006, NewA, 11, 567

\bibitem[\protect\citeauthoryear{Mor, Netzer, \&
    Elitzur}{2009}]{mor09} Mor R., Netzer H., Elitzur M., 2009, ApJ,
  705, 298

\bibitem[\protect\citeauthoryear{}{2008b}]{nenkova08b}
  Nenkova M., Sirocky M.~M., Nikutta R., Ivezi{\'c} {\v Z}., Elitzur
  M., 2008b, ApJ, 685, 160

\bibitem[\protect\citeauthoryear{Nenkova et al.}{2008a, 2008b}]{nenkova08a}
  Nenkova M., Sirocky M.~M., Ivezi{\'c} {\v Z}., Elitzur M., 2008a,
  ApJ, 685, 147

\bibitem[\protect\citeauthoryear{Neugebauer et
    al.}{1979}]{neugebauer79} Neugebauer G., Oke J.~B., Becklin E.~E.,
  Matthews K., 1979, ApJ, 230, 79

\bibitem[\protect\citeauthoryear{Peeters, Spoon, \&
    Tielens}{2004}]{peeters04} Peeters E., Spoon H.~W.~W., Tielens
  A.~G.~G.~M., 2004, ApJ, 613, 986

\bibitem[\protect\citeauthoryear{Piconcelli et
    al.}{2005}]{piconcelli05} Piconcelli E., Jimenez-Bail{\'o}n E.,
  Guainazzi M., Schartel N., Rodr{\'{\i}}guez-Pascual P.~M.,
  Santos-Lle{\'o} M., 2005, A\&A, 432, 15

\bibitem[\protect\citeauthoryear{Pier \& Krolik}{1992}]{pier92} Pier
  E.~A., Krolik J.~H., 1992, ApJ, 401, 99

\bibitem[\protect\citeauthoryear{Pineau et al.}{2011}]{pineau11}
  Pineau F.-X., Motch C., Carrera F., Della Ceca R., Derri{\`e}re S.,
  Michel L., Schwope A., Watson M.~G., 2011, A\&A, 527, A126

\bibitem[\protect\citeauthoryear{Porquet et al.}{2004}]{porquet04}
  Porquet D., Reeves J.~N., O'Brien P., Brinkmann W., 2004, A\&A, 422,
  85

\bibitem[\protect\citeauthoryear{Press et al.}{1992}]{press92} 
Press W.~H., Teukolsky S.~A., Vetterling W.~T., Flannery B.~P., 1992, 
nrfa.book,  

\bibitem[\protect\citeauthoryear{Ramos Almeida et al.}{2011}]{ramos11}
  Ramos Almeida C., et al., 2011, ApJ, 731, 92

\bibitem[\protect\citeauthoryear{Ramos Almeida et al.}{2007}]{ramos07}
  Ramos Almeida C., P{\'e}rez Garc{\'{\i}}a A.~M., Acosta-Pulido
  J.~A., Rodr{\'{\i}}guez Espinosa J.~M., 2007, AJ, 134, 2006

\bibitem[\protect\citeauthoryear{Richards et al.}{2006}]{richards06}
  Richards G.~T., et al., 2006, AJ, 131, 2766

\bibitem[\protect\citeauthoryear{Roseboom et al.}{2013}]{roseboom13}
  Roseboom I.~G., Lawrence A., Elvis M., Petty S., Shen Y., Hao H.,
  2013, MNRAS, 429, 1494

\bibitem[\protect\citeauthoryear{Rovilos et al.}{2014}]{rovilos14}
  Rovilos E., et al., 2014, MNRAS, 438, 494

\bibitem[\protect\citeauthoryear{Scott et al.}{2011}]{scott11} Scott
  A.~E., Stewart G.~C., Mateos S., Alexander D.~M., Hutton S., Ward
  M.~J., 2011, MNRAS, 417, 992

\bibitem[\protect\citeauthoryear{Scott, Stewart, \&
    Mateos}{2012}]{scott12} Scott A.~E., Stewart G.~C., Mateos S.,
  2012, MNRAS, 423, 2633

\bibitem[\protect\citeauthoryear{Silva, Maiolino, \&
    Granato}{2004}]{silva04} Silva L., Maiolino R., Granato G.~L.,
  2004, MNRAS, 355, 973

\bibitem[\protect\citeauthoryear{Simpson}{2005}]{simpson05} Simpson
  C., 2005, MNRAS, 360, 565

\bibitem[\protect\citeauthoryear{Stern et al.}{2012}]{stern12} Stern
  D., et al., 2012, ApJ, 753, 30

\bibitem[\protect\citeauthoryear{Tozzi et al.}{2006}]{tozzi06} Tozzi
  P., et al., 2006, A\&A, 451, 457

\bibitem[\protect\citeauthoryear{Treister, Krolik, \&
    Dullemond}{2008}]{treister08} Treister E., Krolik J.~H., Dullemond
  C., 2008, ApJ, 679, 140

\bibitem[\protect\citeauthoryear{Turner et al.}{1997}]{turner97}
  Turner T.~J., George I.~M., Nandra K., Mushotzky R.~F., 1997, ApJS,
  113, 23

\bibitem[\protect\citeauthoryear{Urry \& Padovani}{1995}]{urry95} Urry
  C.~M., Padovani P., 1995, PASP, 107, 803

\bibitem[\protect\citeauthoryear{Vasudevan \&
    Fabian}{2007}]{vasudevan07} Vasudevan R.~V., Fabian A.~C., 2007,
  MNRAS, 381, 1235

\bibitem[\protect\citeauthoryear{Winter et al.}{2012}]{winter12}
  Winter L.~M., Veilleux S., McKernan B., Kallman T.~R., 2012, ApJ,
  745, 107

\bibitem[\protect\citeauthoryear{Winter et al.}{2009}]{winter09}
  Winter L.~M., Mushotzky R.~F., Reynolds C.~S., Tueller J., 2009,
  ApJ, 690, 1322

\bibitem[\protect\citeauthoryear{Wright et al.}{2010}]{wright10} Wright E.~L.,
  Eisenhardt P.~R.~M., Mainzer A.~K., et al.\ 2010, \aj, 140, 1868

\bibitem[\protect\citeauthoryear{Yan et al.}{2013}]{yan13} Yan L., et
  al., 2013, AJ, 145, 55

\bibitem[\protect\citeauthoryear{Yang, Wang, \& Liu}{2014}]{yang14}
  Yang H., Wang J., Liu T., 2014, arXiv, arXiv:1411.4585

\end{thebibliography}
\end{document}